\documentclass[longauth]{aa}
\usepackage{graphicx}
\usepackage{txfonts}

\usepackage{natbib}
\bibpunct{(}{)}{;}{a}{}{,}
\usepackage{epsf}
\usepackage{lscape}

\newcommand{\bdv}[1]{\mbox{\boldmath$#1$}}
\def\bpi{{\bdv{\pi}}}

\def\bmu{{\bdv{\mu}}}
\def\rel{{\rm{rel}}}

\begin{document}

\title{MOA-2009-BLG-387Lb: A massive planet orbiting an M dwarf}
\titlerunning{MOA-2009-BLG-387}
\authorrunning{V.~Batista \emph{et al.}}

\author{Virginie Batista\inst{1,2}, A.~Gould\inst{3,4}, S.~Dieters\inst{1,2}, Subo Dong\inst{3,5,6}, I.~Bond\inst{7,8}, J.P.~Beaulieu\inst{1,2}, D.~Maoz\inst{3,9}, B.~Monard\inst{3,10}, G.W.~Christie\inst{3,11}, J.~McCormick\inst{3,12}, M.D.~Albrow\inst{1,13},  K.~Horne\inst{1,14,15}, ,Y.~Tsapras\inst{1,14,66,67}, M.J.~Burgdorf\inst{16,62,63}, S.~Calchi Novati\inst{16,69,71,70}, J.~Skottfelt\inst{16,17},  J.~Caldwell\inst{1,19}, S.~Koz{\l}owski\inst{4}, D.~Kubas\inst{1,2,20}, B.S.~Gaudi\inst{3,4}, C.~Han\inst{3,21},  D.~P.~Bennett\inst{1,7,22}, J.~An\inst{81}\\
and\\
F.~Abe\inst{23}, C.S.~Botzler\inst{24}, D.~Douchin\inst{24}, M.~Freeman\inst{24}, A.~Fukui\inst{23}, K.~Furusawa\inst{23}, J.B.~Hearnshaw\inst{13}, S.~Hosaka\inst{23}, Y.~Itow\inst{23}, K.~Kamiya\inst{23}, P.M.~Kilmartin\inst{25}, A.~Korpela\inst{26}, W.~Lin\inst{8}, C.H.~Ling\inst{8}, S.~Makita\inst{23}, K.~Masuda\inst{23}, Y.~Matsubara\inst{23}, N.~Miyake\inst{23}, Y.~Muraki\inst{27}, M.~Nagaya\inst{23}, K.~Nishimoto\inst{23}, K.~Ohnishi\inst{28}, T.~Okumura\inst{23}, Y.C.~Perrott\inst{24}, N.~Rattenbury\inst{24,14}, To.~Saito\inst{29}, D.J.~Sullivan\inst{26}, T.~Sumi\inst{23}, W.L.~Sweatman\inst{8}, P.J.~Tristram\inst{25}, E.~von Seggern\inst{24}, P.C.M.~Yock\inst{24}\\
({The MOA Collaboration}),\\
and\\
S.~Brillant\inst{20}, J.J.~Calitz\inst{30}, A.~Cassan\inst{2}, A.~Cole\inst{31}, K.~Cook\inst{32}, C.~Coutures\inst{33}, D.~Dominis Prester\inst{34}, J.~Donatowicz\inst{35}, J.~Greenhill\inst{31}, M.~Hoffman\inst{30}, F.~Jablonski\inst{37}, S.R.~Kane\inst{38}, N.~Kains\inst{14,15,61}, J.-B.~Marquette\inst{2}, R.~Martin\inst{39}, E.~Martioli\inst{37}, P.~Meintjes\inst{30}, J.~Menzies\inst{40}, E.~Pedretti\inst{15}, K.~Pollard\inst{13}, K.C.~Sahu\inst{41}, C.~Vinter\inst{17}, J.~Wambsganss\inst{42,16}, R.~Watson\inst{31}, A.~Williams\inst{39,82}, M. Zub\inst{42,43}\\
({The PLANET Collaboration}),\\
and\\
W.~Allen\inst{44}, G.~Bolt\inst{45}, M.~Bos\inst{46}, D.L.~DePoy\inst{47}, J.~Drummond\inst{48}, J.D.~Eastman\inst{4}, A.~Gal-Yam\inst{49}, E.~Gorbikov\inst{9,51}, D.~Higgins\inst{50}, J.~Janczak\inst{4}, S.~Kaspi\inst{9,51}, C.-U.~Lee\inst{52}, F.~Mallia\inst{53}, A.~Maury\inst{53}, L.A.G.~Monard\inst{10}, D.~Moorhouse\inst{54}, N.~Morgan\inst{4}, T.~Natusch\inst{55}, E.O.~Ofek\inst{56,57}, B.-G.~Park\inst{52}, R.W.~Pogge\inst{4}, D.~Polishook\inst{9}, R.~Santallo\inst{58}, A.~Shporer\inst{9}, O.~Spector\inst{9}, G.~Thornley\inst{54}, J.C.~Yee\inst{4}\\
({The $\mu$FUN Collaboration}),\\
and\\
V.~Bozza\inst{69,71,70}, P.~Browne\inst{15}, M.~Dominik\inst{15,72}, S.~Dreizler\inst{73}, F.~Finet\inst{74}, M.~Glitrup\inst{75}, F.~Grundahl\inst{75}, K.~Harps{\o}e\inst{17}, F.V.~Hessman\inst{73}, T.C.~Hinse\inst{17,76}, M.~Hundertmark\inst{73}, U.G.~J{\o}rgensen\inst{17,18}, C.~Liebig\inst{15,42}, G.~Maier\inst{42}, L.~Mancini\inst{69,70,80}, M.~Mathiasen\inst{17}, S.~Rahvar\inst{77,79}, D.~Ricci\inst{74}, G.~Scarpetta\inst{69,71,70},  J.~Southworth\inst{78}, J.~Surdej\inst{74}, F.~Zimmer\inst{17,42}\\
({The MiNDSTEp Consortium})\\
A.~Allan\inst{59}, D.M.~Bramich\inst{1,61},C.~Snodgrass\inst{16,20,65},I.A.~Steele\inst{60},R.A.~Street\inst{67,68}\\
({The RoboNet Collaboration}),\\
and\\
}

\institute { Probing Lensing Anomalies NETwork (PLANET)
\and Institut d'Astrophysique de Paris, Universit{\'e} Pierre et Marie Curie, CNRS UMR7095, 98bis Boulevard Arago, 75014 Paris, France; batista,beaulieu,cassan,marquett@iap.fr 
\and Microlensing Follow Up Network ($\mu$FUN) 
\and Department of Astronomy, Ohio State University,140 W.\ 18th Ave., Columbus, OH 43210, USA; 
gould,gaudi,jdeast,jyee,pogge,simkoz@astronomy.ohio-state.edu, nick.morgan@alum.mit.edu
\and Institute for Advanced Study, Einstein Drive, Princeton, NJ 08540, USA; dong@ias.edu
\and Sagan Fellow
\and Microlensing Observations in Astrophysics (MOA)
\and Institute of Information and Mathematical Sciences, Massey University, Private Bag 102-904, North Shore Mail Centre, Auckland, New Zealand; i.a.bond,l.skuljan,w.lin,c.h.ling,w.sweatman@massey.ac.nz
\and School of Physics and Astronomy and Wise Observatory, Tel-Aviv University, Tel-Aviv 69978, Israel; shai,dani,david,shporer,odedspec@wise.tau.ac.il
\and Bronberg Observatory, Centre for Backyard Astrophysics, Pretoria, South Africa; lagmonar@nmisa.org
\and Auckland Observatory, Auckland, New Zealand; gwchristie@christie.org.nz
\and Farm Cove Observatory, Centre for Backyard Astrophysics, Pakuranga, Auckland, New Zealand; farmcoveobs@xtra.co.nz
\and  University of Canterbury, Department of Physics and Astronomy, Private Bag 4800, Christchurch 8020, New Zealand; Michael.Albrow@canterbury.ac.nz
\and  The RoboNet Collaboration
\and SUPA School of Physics and Astronomy, Univ. of St Andrews, Scotland KY16 9SS, United Kingdom; md35,nk87,ep41,kdh1@st-andrews.ac.uk
\and Microlensing Network for the Detection of Small Terrestrial Exoplanets (MiNDSTEp)
\and Niels Bohr Institutet, K{\o}benhavns Universitet, Juliane Maries Vej 30, 2100 K{\o}benhavn {\O}, Denmark
\and Centre for Star and Planet Formation, K{\o}benhavns Universitet, {\O}ster Voldgade 5-7, 1350 K{\o}benhavn {\O}, Denmark
\and McDonald Observatory, 16120 St Hwy Spur 78 \#2, Fort Davis, TX 79734 USA; caldwell@astro.as.utexas.edu
\and European Southern Observatory, Casilla 19001, Santiago 19, Chile; sbrillan,dkubas@eso.org
\and Department of Physics, Institute for Basic Science Research, Chungbuk National University, Chongju 361-763, Korea; cheongho@astroph.chungbuk.ac.kr
\and University of Notre Dame, Department of Physics, 225 Nieuwland Science Hall, Notre Dame, IN 46556-5670 USA; bennett@nd.edu
\and Solar-Terrestrial Environment Laboratory, Nagoya University, Nagoya, 464-8601, Japan; sumi, abe, afukui, furusawa, itow, kkamiya, kmasuda, ymatsu, nmiyake, mnagaya, okumurat, sako@stelab.nagoya-u.ac.jp
\and Department of Physics, University of Auckland, Private Bag 92019, Auckland, New Zealand; c.botzler,p.yock@auckland.ac.nz,yper006@aucklanduni.ac.nz
\and Mt.\ John Observatory, P.O. Box 56, Lake Tekapo 8780, New Zealand
\and School of Chemical and Physical Sciences, Victoria University, Wellington, New Zealand; a.korpela@niwa.co.nz,denis.sullivan@vuw.ac.nz
\and Department of Physics, Konan University, Nishiokamoto 8-9-1, Kobe 658-8501, Japan
\and Nagano National College of Technology, Nagano 381-8550, Japan
\and Tokyo Metropolitan College of Industrial Technology, Tokyo 116-8523, Japan
\and University of the Free State, Faculty of Natural and Agricultural Sciences, Department of Physics, PO Box 339, Bloemfontein 9300, South Africa; HoffmaMJ.SCI@mail.uovs.ac.za
\and University of Tasmania, School of Mathematics and Physics, Private Bag 37, GPO, Hobart, Tas 7001, Australia; John.Greenhill,Andrew.Cole@utas.edu.au	
\and Lawrence Livermore National Laboratory, Institute of Geophysics and Planetary Physics, P.O. Box 808, Livermore, CA 94551-0808 USA; kcook@llnl.gov
\and CEA/Saclay, 91191 Gif-sur-Yvette cedex, France; coutures@iap.fr
\and Department of Physics, University of Rijeka, Omladinska 14, 51000 Rijeka, Croatia
\and Technische Universitaet Wien, Wieder Hauptst. 8-10, A-1040 Wienna, Austria; donatowicz@tuwien.ac.at
\and LATT, Universit\'{e} de Toulouse, CNRS, France
\and Instituto Nacional de Pesquisas Espaciais, Sao Jose dos Campos, SP, Brazil
\and NASA Exoplanet Science Institute, Caltech, MS 100-22, 770 south Wilson Avenue, Pasadena, CA 91125, USA; skane@ipac.caltech.edu
\and Perth Observatory, Walnut Road, Bickley, Perth 6076, WA, Australia; Ralph.Martin, Andrew.Williams@dec.wa.gov.au
\and South African Astronomical Observatory, PO box 9, Observatory 7935, South Africa
\and Space Telescope Science Institute, 3700 San Martin Drive, Baltimore, MD 21218, USA
\and Astronomisches Rechen-Institut, Zentrum f\"{u}r Astronomie der Universit\"{a}t Heidelberg (ZAH),  M\"{o}nchhofstr. 12-14, 69120 Heidelberg, Germany
\and Institute of Astronomy, University of Zielona G\'ora, Lubuska st. 2, 65-265 Zielona G\'ora, Poland
\and Vintage Lane Observatory, Blenheim, New Zealand; whallen@xtra.co.nz
\and Perth, Australia; gbolt@iinet.net.au
\and Molehill Astronomical Observatory, Auckland, New Zealand; molehill@ihug.co.nz
\and Department of Physics and Astronomy, Texas A\&M University, College Station, TX, USA; depoy@physics.tamu.edu
\and Possum Observatory, Patutahi, New Zealand; john\_drummond@xtra.co.nz
\and Department of Particle Physics and Astrophysics, Weizmann Institute of Science, 76100 Rehovot,Israel; avishay.gal-yam@weizmann.ac.il
\and Hunters Hill Observatory, Canberra, Australia; dhi67540@bigpond.net.au
\and Department of Physics, Technion, Haifa 32000, Israel
\and Korea Astronomy and Space Science Institute, Daejon 305-348, Korea; leecu,bgpark@kasi.re.kr
\and Campo Catino Austral Observatory, San Pedro de Atacama, Chile; francomallia@campocatinobservatory.org,alain@spaceobs.com
\and Kumeu Observatory, Kumeu, New Zealand; acrux@orcon.net.nz,guy.thornley@gmail.com
\and AUT University, Auckland, New Zealand; tim.natusch@aut.ac.nz
\and Palomar Observatory, California, USA; eran@astro.caltech.edu
\and Einstein Fellow
\and Southern Stars Observatory, Faaa, Tahiti, French Polynesia; obs930@southernstars-observatory.org
\and School of Physics, University of Exeter, Stocker Road, Exeter EX4 4QL, United Kingdom
\and Astrophysics Research Institute, Liverpool John Moores University, Liverpool CH41 1LD, United Kingdom
\and European Southern Observatory, Karl-Schwarzschild-Stra{\ss}e 2, 85748 Garching bei M\"{u}nchen, Germany
\and Deutsches SOFIA Institut, Universit\"{a}t Stuttgart, Pfaffenwaldring 31, 70569 Stuttgart, Germany
\and SOFIA Science Center, NASA Ames Research Center, Mail Stop N211-3, Moffett Field CA 94035, USA
\and Jodrell Bank Centre for Astrophysics, The University of Manchester, Oxford Road, Manchester M13 9PL, United Kingdom; Eamonn.Kerins@manchester.ac.uk
\and Max Planck Institute for Solar System Research, Max-Planck-Str. 2, 37191 Katlenburg-Lindau, Germany 
\and Astronomy Unit, School of Mathematical Sciences, Queen Mary, University of London, Mile End Road, London, E1 4NS, United Kingdom
\and Las Cumbres Observatory Global Telescope network, 6740 Cortona Drive, suite 102, Goleta, CA 93117, USA
\and Dept. of Physics, Broida Hall, University of California, Santa Barbara CA 93106-9530, USA
\and Universit\`{a} degli Studi di Salerno, Dipartimento di Fisica ''E.R.~Caianiello'', Via Ponte Don Melillo, 84085 Fisciano (SA), Italy
\and Istituto Internazionale per gli Alti Studi Scientifici (IIASS), Via G.\ Pellegrino 19, 84019 Vietri sul Mare (SA), Italy
\and INFN, Gruppo Collegato di Salerno, Sezione di Napoli, Italy
\and Royal Society University Research Fellow
\and Institut f\"{u}r Astrophysik, Georg-August-Universit\"{a}t, Friedrich-Hund-Platz 1, 37077 G\"{o}ttingen, Germany
\and Institut d'Astrophysique et de G\'{e}ophysique, All\'{e}e du 6 Ao\^{u}t 17, Sart Tilman, B\^{a}t.\ B5c, 4000 Li\`{e}ge, Belgium
\and Department of Physics \& Astronomy, Aarhus University, Ny Munkegade 120, 8000 {\AA}rhus C, Denmark
\and Armagh Observatory, College Hill, Armagh, BT61 9DG, United Kingdom 
\and Department of Physics, Sharif University of Technology, P.~O.\ Box 11155--9161, Tehran, Iran 
\and Astrophysics Group, Keele University, Staffordshire, ST5 5BG, United Kingdom
\and School of Astronomy, IPM (Institute for Studies in Theoretical Physics and Mathematics), P.O. Box 19395-5531, Tehran, Iran
\and Dipartimento di Ingegneria, Universit\`{a} del Sannio, Corso Garibaldi 107, 82100 Benevento, Italy
\and National Astronomical Observatories, Chinese Academy of Sciences, A20 Datun Road, Chaoyang District, Beijing 100012, P.R.China; jinan@nao.cas.cn
\and School of Physics, University of Western Australia, Perth, WA 6009}


\date{Accepted} 
\abstract 
{}
{We report the discovery of a planet with a high planet-to-star mass ratio in the
microlensing event MOA-2009-BLG-387, which exhibited pronounced
deviations over a 12-day interval, one of the longest for any planetary
event. The host is an M dwarf, 
with a mass in the range $0.07\,M_\odot < M_{\rm host}< 0.49\,M_\odot$ 
at 90\% confidence.
The planet-star mass ratio $q=0.0132\pm 0.003$ 
has been measured extremely well, so at the best-estimated host mass, the
planet mass is $m_p = 2.6$ Jupiter masses for the median host mass,
$M=0.19\,M_\odot$.} {The host mass is determined
from two ``higher order'' microlensing parameters.  One of these,
the angular Einstein radius $\theta_{\rm E}=0.31\pm 0.03\,$mas has been accurately measured, but the other (the microlens parallax $\pi_{\rm E}$,
which is due to the Earth's orbital motion) is highly degenerate with
the orbital motion of the planet. We statistically resolve the degeneracy
between Earth and planet orbital effects by imposing priors
from a Galactic model that specifies the positions and velocities
of lenses and sources and a Kepler model of orbits.}{ The 90\%
confidence intervals for the distance, semi-major axis, and period
of the planet are $3.5\,{\rm kpc} < D_L < 7.9\,{\rm kpc}$,
$1.1\,{\rm AU}< a < 2.7\,{\rm AU}$, and 
$3.8\,{\rm yr} < P < 7.6\,{\rm yr}$, respectively.}
{}


\maketitle


\section{Introduction}

Over the past decade, the gravitational microlensing method has led to
detection of ten exoplanets 
\citep{ob03235,ob05071,ob05390,ob05169,ob06109,mb07192,mb07400,mb08310,ob07368},
which permits the exploration of host-star 
and planet populations whose mass and distance are not probed
by any other method. Indeed, since the efficiency of the microlensing
method does not depend on detecting light from the host star, it
allows one to probe essentially all stellar types over distant 
regions of our Galaxy. In particular, microlensing is an excellent
method to explore planets around M dwarfs, which are the most common stars
in our Galaxy, but which are often a challenge for other techniques
because of their low luminosity.
Roughly half of all microlensing events toward the Galactic bulge stem from stars with mass $\la 0.5~M_\odot$ \citep{gould00a}.

Determining the characteristics and frequency of planets orbiting M
dwarfs is of interest not only because M dwarfs are the most common
type of stars in the Galaxy, but also because these systems provide
important tests of planet formation theories.  In particular, the core
accretion theory of giant planet formation predicts that giant planets
should be less common around low-mass stars
\citep{laughlin04,ida05,kennedy08,dangelo10}, whereas the
gravitational instability model predicts that giant planets can form
around M dwarfs with sufficiently massive protoplanetary disks
\citep{boss06}.  In fact, there is accumulating evidence from radial
velocity surveys that giant planets are less common around low-mass
primaries \citep{cumming08,johnson10}.  However, these surveys are only
sensitive to planets with semimajor axes of $< 2.5~{\rm AU}$.  Since
it is thought that the majority of the giant planets found by radial
velocity surveys likely formed farther out in their protoplanetary
disks and subsequently migrated close to their parent star, it is not
clear whether the relative paucity of giant planets around low-mass stars
found in these surveys is a statement about the dependence on
stellar mass of migration or of formation.

Microlensing is complementary to the radial velocity technique in that it is
sensitive to planets with larger semimajor axes, closer to their
supposed birth sites. Indeed, based on the analysis of 13 well-monitored
high-magnification events with 6 detected planets, \citet{gould10}
found that the frequency of giant planets at separations of $\sim
2.5~{\rm AU}$ orbiting $\sim 0.5~M_\odot$ hosts was quite high and, in
particular, consistent with the extrapolation of the frequencies of
small-separation giant planets orbiting solar mass hosts inferred from
radial velocity surveys out to the separations where microlensing is 
most sensitive.
This suggests that low-mass stars may form giant planets as efficiently as do
higher mass stars, but that these planets do not migrate as
efficiently.

Furthermore, of the ten previously published microlensing planets, one
was a ``supermassive'' planet with a very high mass ratio: a $m_p = 3.8
M_{\rm Jup}$ planet orbiting an M dwarf of mass
$M = 0.46~M_\odot$ \citep{dong09}. 
Given their high planet-to-star mass ratios $q$, such
planets are expected to be exceedingly rare in the core-accretion
paradigm, so the mere existence of this planet may pose a
challenge to such theories. Gravitational instability, on the other
hand, favors the formation of massive planets (provided they form at
all).

Current and future microlensing surveys are particularly sensitive to
large $q$ planets orbiting M dwarf hosts, for several reasons. As
with other techniques, microlensing is more sensitive to planets with
higher $q$. In addition, as the mass ratio increases, a larger
fraction of systems induce an important subclass of resonant-caustic
lenses. Resonant caustics are created when the planet happens to have
a projected separation close to the Einstein radius of the
primary \citep{wambs97}. The range of separations that give rise
to resonant caustics is quite narrow for small $q$, but grows as
$q^{1/3}$. Furthermore, although the range of parameter space giving
rise to resonant caustics is narrow, the caustics themselves and their
cross sections are large and also grow as $q^{1/3}$.  Thus the
probability of detecting planets via these caustics is relatively
high, and such systems contribute a significant fraction of all
detected events, particularly for supermassive planets orbiting M
dwarfs.  Events due to resonant caustics are particularly valuable, as
they allow one to further constrain the properties and orbit of the planet.
This is because these events usually exhibit caustic features that are
separated well in time.  When combined with the fact that the precise
shape of a resonant caustic is extremely sensitive to the
separation of the planet from the Einstein ring, such light curves are
particularly sensitive to orbital motion of the planet (see, e.g.,
\citealt{bennett10}).

Here we present the analysis of the microlensing event
MOA-2009-BLG-387, a resonant-caustic event, which we demonstrate is caused by a massive planet orbiting an M dwarf.  The light curve associated
with this event contains very prominent caustic features that are well
separated in time. These structures were very intensively monitored by the
microlensing observers, so that the geometry of the system is quite
well constrained. As a result, the event has high sensitivity to two
higher order effects: parallax and orbital motion of the planet. In
Section 4, we present the modeling of these two effects and our
estimates of the event characteristics.  This analysis reveals a degeneracy between one component of the parallax and
one component of the orbital motion. We explain, for the first time, the
causes of this degeneracy.  It gives rise to very
large errors in both the parallax and orbital motion, which makes the
final results highly sensitive to the adopted priors.
In particular,
uniform priors in microlensing variables imply essentially
uniform priors in lens-source relative parallax, whereas the
proper prior for physical location is uniformity in volume element.
These differ by approximately a factor $D_l^4$, where $D_l$ is the
lens distance.  In Section 5, we therefore give a careful
Bayesian analysis that properly weights the distribution by
correct physical priors.  The high-mass end of the range still permitted is eliminated by the failure to detect
flux from the lens using high-resolution NACO images on the VLT.
Combining all available information, we find that the host
is an M dwarf in the mass range
$0.07\,M_\odot < M_{\rm host}< 0.49\,M_\odot$ at 90\% confidence.

\section{Observational data}

The microlensing event MOA-2009-BLG-387 was alerted by the MOA
collaboration (Microlensing Observations in Astrophysics) on 24
July 2009 at 15:08 UT, $HJD'\equiv HJD - 2,450,000 = 5037.13$, a
few days before the first caustic entry. Many observatories obtained data of the event. The celestial coordinates of
the event are $\alpha = 17h53m50.79s$ and $\delta =
-33^{\circ}59'25''$ (J2000.0) corresponding to Galactic coordinates :
$l = +356.56$, $b = -4.097$.

The lightcurve is overall characterized by two pairs of caustic
crossings (entrance plus exit), which together span 12 days (see
Figure\ref{fig:lc}). This structure is caused by the source passing
over two ``prongs'' of a resonant caustic (see Figure\ref{fig:lc}
inset). Obtaining good coverage of these caustic crossings posed a
variety of challenges.

The first caustic entrance ($HJD' = 5040.3$) was detected by the
PLANET collaboration using the South African Astronomical Observatory
(SAAO) at Sutherland (Elizabeth 1m) who then issued an anomaly alert
at $HJD'$ 5040.4 calling for intensive follow-up observations, which in turn
enabled excellent coverage of the first caustic exit roughly one day
later.

The second caustic entrance occurred about seven days later ($HJD' =
5047.1$, see Figure\ref{fig:lc}). That the caustic crossings
are so far apart in time is quite unusual in planetary microlensing
events. Since round-the-clock intensive observations cannot normally
be sustained for a week, accurate real-time prediction of the second caustic
entrance was important for obtaining intensive coverage of
this feature. In fact, the second caustic entrance was predicted 14
hours in advance, with a five-hour discrete uncertainty due to the well-known
close/wide $s\leftrightarrow s^{-1}$ degeneracy, where $s$ is the
projected separation in units of the Einstein radius. The close-geometry
crossing prediction was accurate to less than one half hour and the
caustic-geometry prediction was almost identical to the one derived
from the best fit to the full lightcurve, which is shown in
Figure\ref{fig:lc}.

The extended duration of the lightcurve anomalies indicates a
correspondingly large caustic structure. Indeed, the preliminary models
found a planet/star separation (in units of Einstein
radius) close to unity, 
which means that the caustic is resonant (see the caustic
shape in the upper panel of Figure\ref{fig:lc}, where the source is
going upward).
 
The event was alerted and monitored by the MOA collaboration. It was
also monitored by the Probing Lensing Anomalies Network collaboration
(PLANET; \citealt{albrow98}) from three different telescopes: at the
South African Astronomical Observatory (SAAO), as mentioned above, as
well as the Canopus 1 m at Hobart (Tasmania) and the 60 cm of Perth
Observatory (Australia).

The Microlensing Follow Up Network ($\mu$FUN ; \citealt{yoo04})
followed the event from Chile (1.3m SMARTS telescope at CTIO)
(\textit{V}, \textit{I} and \textit{H} band data), 
South Africa (0.35 m telescope
at Bronberg observatory), New Zealand (0.40 m and 0.35 m telescopes at
Auckland Observatory (AO) and Farm Cove (FCO) observatory,
respectively, the Wise observatory (1.0 m at Mitzpe Ramon, Israel), and the Kumeu observatory (0.36 m telescope at Auckland, NZ).

The RoboNet collaboration also followed the event with their three 2m
robotic telescopes : the Faulkes Telescopes North (FTN) and South
(FTS) in Hawaii and Australia (Siding Springs Observatory)
respectively, and the Liverpool Telescope (LT) on La Palma (Canary
Islands). And finally, the MiNDSTEp collaboration observed the event
with the Danish 1.54 m at ESO La Silla (Chile).

 Observational
conditions for this event were unusually challenging, due in part to
the faintness of the target and the presence of a bright neighboring
star. Moreover, the full moon passed close to the source near the
second caustic entrance. As a result, several data sets were of much
lower statistical quality and had much stronger systematics than the
others. We therefore selected seven data sets that cover the caustic
features and the entire lightcurve : MOA, SAAO, FCO, AO, Danish,
Bronberg, and Wise. They include 118 MOA data points in \textit{I}
band, 221 PLANET data points in \textit{I} band, 262 $\mu$FUN data
points in unfiltered, \textit{R} and \textit{I} bands, and 300 MiNDSTEp
data points in \textit{I} band. We also fit the $\mu$FUN CTIO \textit{I}
and \textit{V} data to the final model, but solely for the purpose of
determining the source size.  And finally, we fit $\mu$FUN CTIO 
$H$-band data to the lightcurve in order to compare the $H$-band
source flux with the late-time $H$-band baseline flux from VLT images
(see Section~\ref{sec:VLT}).
The SAAO, FCO, AO, Danish, Bronberg, and Wise data were reduced by MDA using the PYSIS3 software \citep{albrow09}. The FCO, AO, Bronberg, and Wise images were taken in white light and suffered from systematic
effects related to the airmass. Such effects were corrected by extracting lightcurves of other stars in the field with similar colors
to the lens, and assuming that these stars are intrinsically constant.

For each data set, the errors were
rescaled to make $\chi^2$ per degree of freedom for the best
binary-lens fit close to unity. We then eliminated the largest outlier
and repeated the process until there were no 3 $\sigma$ outliers.

{\subsection{VLT NACO Images}
\label{sec:VLT}}

On 7 June 2010, we obtained high-resolution $H$-band images using the
NACO imager on the Very Large Telescope (VLT). Since this was approximately
7.7 Einstein timescales after the peak of the event, the source was essentially
at the baseline. The reduction procedures were similar to those of
MOA-2008-BLG-310, which are described in detail by \citet{mb08310}.

To identify the source on the NACO frame, we first performed image 
subtraction on CTIO $I$-band images to locate its position on
the $I$-band frame.  We then used the NACO image to find relatively
unblended stars that could be used to align the $I$-band and NACO
frames.  There is clearly a source at the inferred position, but
it lies only seven pixels ($0.19''$) from an ambient star, which
is 1.35 mag brighter than the ``target'' (source plus lens plus
any other blended light within the aperture).  This proximity
induces a 94\% correlation coefficient between the photometric
measurements of the two stars.  We therefore estimate the target
error as 0.06 mag. In the NACO system (which is calibrated to 2MASS
using comparison stars) the target magnitude is
\begin{equation}
H_{\rm target,NACO} = 18.25 \pm 0.06 .
\label{eqn:htarget}
\end{equation}

We have an $H$-band light curve (taken simultaneously with $V$ and $I$
at CTIO), and so (once we have established a model fit the light curve
in Section~{\ref{sec:model}) we can measure quite precisely the source 
flux in the
CTIO system, $H_{\rm source,CTIO} = 20.03\pm 0.02$.  To compare
with NACO, we transform to the NACO system using 4 comparison stars
that are relatively unblended, a process to which we assign a 0.03 mag
error, finding
\begin{equation}
H_{\rm source,NACO} = 18.35 \pm 0.03.
\label{eqn:hsource}
\end{equation}
The difference, consisting of light from the lens as well as any other
blended light in the aperture, is $0.10\pm 0.07$.

This excess-flux measurement could in principle be due to five
physical effects.  First, it is reasonably consistent with normal
statistical noise.  Second, it could come from the lens.  As we
show in Section~\ref{sec:bayesian}, this would be consistent with
a broad range of M dwarf lenses.  Third, it could be a companion
to the source, and fourth, a companion to the lens.  Finally,
it could be an ambient star unrelated to the event.  The fundamental
importance of this measurement is that, for all five of these 
possibilities, the measurement places an upper limit on the flux
from the lens, hence its mass (assuming it is not a white dwarf).

\section{Source properties from color-magnitude diagram and measurement of $\theta_{\rm E}$}
\label{cmd}

To determine the dereddened color and magnitude of the microlensed
source, we put the best fit color and magnitude of the source on an
$(I,V-I)$ instrumental color magnitude diagram (CMD)
(cf. Fig.\ref{fig:cmd}), using instrumental CTIO data.  The magnitude
and color of the target are $I=20.62\pm 0.04$ and $(V-I)=-0.42\pm
0.01$. The mean position of the red clump is represented by an open
circle at $(I,V-I)_{RC}=(16.36,-0.16)$, with an error of 0.05 for both
quantities.

For the absolute clump magnitude, we adopt $M_{I,RC}=-0.25 \pm 0.05$ 
from \citet{bennett10}. We adopt
the measured bulge clump color $(V-I)_{0,RC}=1.08 \pm 0.05$ 
(Fig.~5 of \citealt{bensby10}) and a Galactocentric distance
$R_0 = 8.0\pm 0.3\,$kpc \citep{yelda10}.  We further assume that at
the longitude ($l=-3.4$), the bar lies 0.7 kpc more distant than $R_0$
(D. Nataf et al., in preparation), i.e., 8.7 kpc.  From this, we derive
$(I,V-I)_{0,RC}=(14.45,1.08)\pm(0.10,0.05)$, so that the dereddened
source color and magnitude are given by:
$(I,V-I)_0=\Delta(I,V-I)+(I,V-I)_{0,RC}=(18.71,0.82)$.
From $(V-I)_0$, we derive $(V-K)_0=1.78\pm 0.14$ using the \cite{bessbret88}
color-color relations.

The color determines the relation between dereddened source flux and
angular source radius,
\citep{kerv04} 
\begin{equation}
\log{2\theta_*}=0.5170-0.2V_0+0.2755(V-K)_0,
\label{eqn:source}
\end{equation}
giving $\,\theta_*=0.63\pm 0.06\,\mu {\rm as}$.
With the angular size of the source given by the limb-darkened extended-source 
 fit (model 5, see Table~1), 
$\rho_*=0.00202\pm0.00003$, we derive the angular Einstein radius
$\theta_{\rm E}$ : $\theta_{\rm E}=\theta_*/\rho_*=0.31\pm 0.03\,\rm{mas}$.

\section{Event modeling}
\label{sec:model}
\subsection{Overview}

The modeling proceeds in several stages. We first give an overview of
these stages and then consider them each in detail. First, inspection
of the lightcurve shows that the source crossed over two ``prongs'' of
a caustic, or possibly two separate caustics, with a pronounced
trough in between.
The source spent 1-3 days crossing each prong and 7 days
between prongs. This pattern strongly implies that the event
topology is that of a source crossing the ``back end'' of a
resonant caustic with $s<1$, as illustrated in
Figure\ref{fig:lc}. We nevertheless conducted a blind search of
parameter space, incorporating the minimal 6 standard static-binary
parameters required to describe all binary events, as well as
$\rho=\theta_*/\theta_{\rm E}$, the source size in units of the
Einstein radius.  The parameters derived from this fit are quite
robust. However, they yield only the planet-star mass ratio $q$, but
not the planet mass $m_p=q M$, where $M$ is the host mass. In
principle, one can measure $M$ from (e.g. \citealt{gould00b})
\begin{equation}
M=\frac{\theta_{\rm E}}{\kappa \pi_{\rm E}}
\label{eqn:mval}
\end{equation}
where $\pi_{\rm E}$ is the ``microlens parallax'' and $\kappa\equiv
4G/(c^2\,{\rm AU})\sim 8.1\,{\rm mas}\, M_\odot^{-1}$. However, while
$\theta_{\rm E}=\theta_* / \rho$ is also quite robustly determined
from the static solution (and Section~\ref{cmd}), $\pi_{\rm E}$ is not.

However, the event timescale is moderately long  ($\sim 40$ days).
This would not normally be long enough to
measure the full microlens parallax, but might be enough to
measure one dimension of the parallax vector \citep{gouldmiralda94}.
Moreover, the large
separation in time of the caustic features could permit detection of
orbital motion effects as well \citep{mb9741}.
We therefore incorporate these two
effects, first separately and then together. We find that each is
separately detected with high significance, but that when combined
they are partially degenerate with each other. In particular, one of
the two components of the microlensing parallax vector $\vec\pi_{\rm
E}$ is highly degenerate with one of the two measurable parameters of
orbital motion. It is often the case that one or both components of
$\vec\pi_{\rm E}$ are poorly measured in planetary microlensing
events. The usual solution is to adopt Bayesian priors for the
lens-source relative parallax and proper motion, based on a Galactic
model. We also pursue this approach, but in addition we consider
separately Bayesian priors on the orbital parameters as well. We show
that the results obtained by employing either set of priors separately
are consistent with each other, and we therefore combine both sets of
priors.

\onecolumn{
\begin{figure}[h!]
\centering
\includegraphics[width=4.5in]{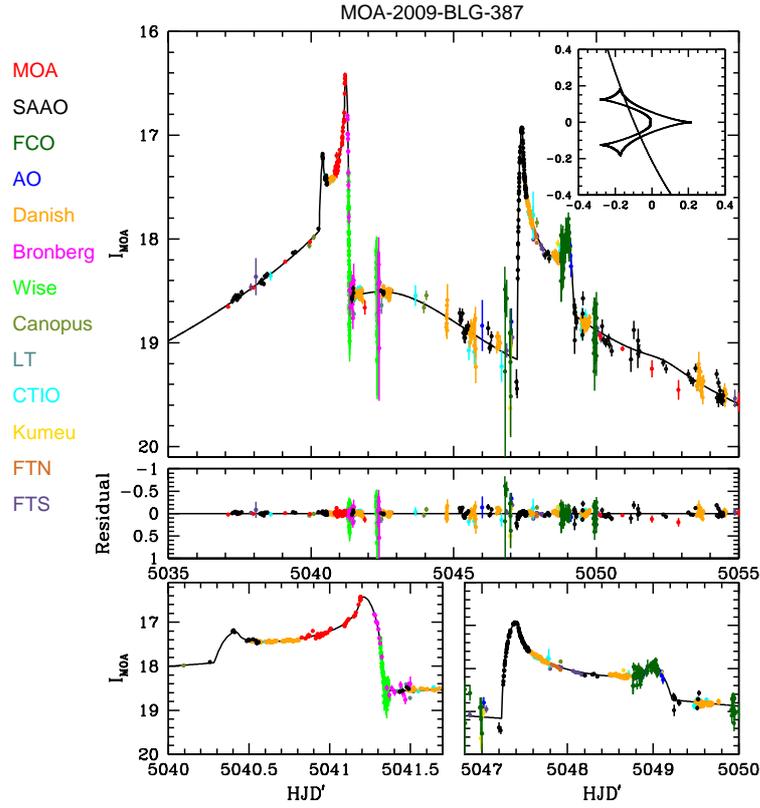}
\caption{\label{fig:lc}
Top: Light curve of MOA-2009-BLG-387 near its peak in July 2009 and
the trajectory of the source across the caustic feature on the
right. The source is going upward. We show the model with
finite-source, parallax and orbital motion effects.  Middle:
Magnitude residuals.  Bottom: Zooms of the caustic features of the
light curve.  }
\end{figure}

\begin{figure}[h!]
\centering
\includegraphics[angle=0,width=3.5in]{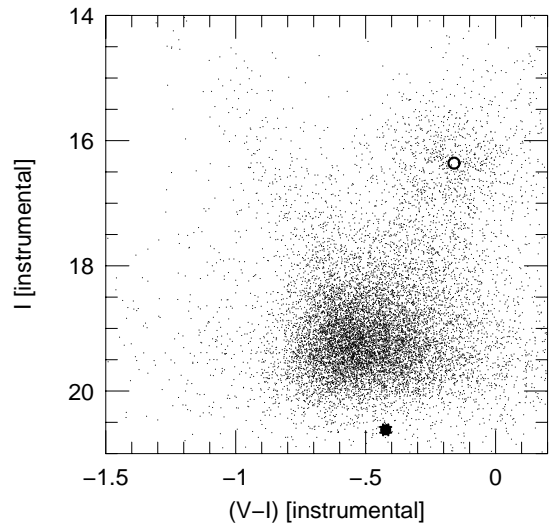}
\caption{\label{fig:cmd}
Instrumental color-magnitude diagram of the field around
MOA-2009-BLG-387. The clump centroid is shown by an open circle,
while the CTIO $I$ and $V-I$ measurements of the source are shown by
a filled circle}
\end{figure}
}

\twocolumn{

\subsection{Static binary}
\label{static}

A static binary-lens point-source model involves six microlensing
parameters: three related to the lens-source kinematics ($t_0$,
$u_0$, $t_E$), where $t_0$ is the time of lens-source closest
approach, $u_0$ is the impact parameter with respect to the center of
mass of the binary-lens system and $t_E$ is the Einstein timescale of
the event, and three related to the binary-lens system ($q$, $s$,
$\alpha$), where $q$ and $s$ are the planet-star mass
ratio and separation in units of Einstein radius, respectively, and $\alpha$ is the
angle between the trajectory of the source and the star-planet
axis. For $n=7$ observatories, there are $2n$ photometric parameters,
$n\times (F_s,F_b)$, which correspond to the source flux and blend
flux for each data set. These are usually determined by linear
regression.  The radius of the source, $\rho$, in Einstein units, can
also be derived from the model provided that the source passes over,
or sufficiently close to, a caustic structure.
To optimize the fit in terms of computing time, we adopt
different methods for implementing finite-source effects, depending on
the distance between the source and the caustic features in the sky
plane. When the source is far from the caustic (in the wings of the
lightcurve), we treat it as a point source. In the caustic
crossing regions, we use a finite-source model based on the Green-Stokes
theorem \citep{gouldgauch97}. Numerical implementation of this method is adapted from the code that was originally devised for \cite{albrow01} and refined in \cite{an02}. This technique,
which reduces the 2-dimensional integral over the source to a 1-dimensional 
integral over its boundary and so is extremely efficient, 
implicitly assumes that the source has
uniform surface brightness, i.e., is not limb darkened.  We then
include limb-darkening in the final fit, as described in Section \ref{ld}.
Lastly, in the intermediary regions, we use the
hexadecapole approximation \citep{pejcha09,gould08}, which consists of
calculating the magnification of 13 points distributed over the source
in a characteristic pattern.  
To fit the microlensing parameters, we
perform a Markov Chain Monte Carlo (MCMC) fitting with an adaptive
step-size Gaussian sampler \citep{doran04,dong09}. After every 200
links in the chain, the covariance matrix between the MCMC
parameters is calculated again. We
proceed to five runs corresponding to five different configurations:
without either parallax or orbital motion, with parallax only, with orbital
motion only, with both effects, and finally with both effects and
limb-darkening effects included. The results are presented in
Section~\ref{results1}.

The static binary search without parallax leads to the following
parameters: $q=0.0107$, $s=0.9152$, $\rho=0.00149$, and then
$\theta_E=0.42\,$mas, implying
\begin{equation}
M \pi_{\rm{rel}}=\frac{\theta^2_E}{\kappa}=22\, M_\odot\,\mu{\rm{as}}
\end{equation}
This product is consistent, for example, with a $1\,M_\odot$ mass host
in the Galactic bulge or a $0.025\, M_\odot$ mass brown-dwarf star at
1 kpc, either of which would have very important implications for the
nature of the $q=0.0107$ planet. We therefore first investigate
whether the microlens parallax can be measured.

\subsection{Parallax effects}
\label{parallax}

When observing a microlensing event, the resulting flux for each
observatory-filter $i$ can be expressed as,
\begin{equation}
F_i(t)=F_{s,i}A[u(t)]+F_{b,i},
\label{eqn:flux}
\end{equation}
where $F_{s,i}$ is the flux of the unmagnified source, $F_{b,i}$ is
the background flux and $u(t)$ is the source-lens projected separation
in the lens plane.
The source-lens projected separation in the lens plane, $u(t)$ of
Eq.~(\ref{eqn:flux}), can be expressed as a combination of two
components, $\tau(t)$ and $\beta(t)$, its projections along the
direction of lens-source motion and perpendicular to it, respectively:
\begin{equation}
u(t)=\sqrt{\tau^2(t)+\beta^2(t)}.
\label{eqn:ut}
\end{equation}
If the motion of the source, lens and observer can all be considered
rectilinear, the two components of $u(t)$ are given by
\begin{equation}
\tau(t)=\frac{t-t_0}{t_{\rm E}}\qquad ; \qquad \beta(t)=u_0.
\label{eqn:u1u2}
\end{equation}

To introduce parallax effects, we use the geocentric formalism
\citep{an02,gould04} which ensures that the three standard
microlensing parameters ($t_0$, $t_{\rm E}$, $u_0$) are nearly the
same as for the no-parallax fit. Hence, two more parameters are fitted
in the MCMC code, i.e., the two components of the parallax vector,
$\bpi_{\rm E}$, whose magnitude gives the projected Einstein radius,
$\tilde{r}_{\rm E}=AU/\pi_{\rm E}$ and whose direction is that of
lens-source relative motion.
The parallax effects imply additional terms in the Eq.~(\ref{eqn:u1u2})
\begin{equation}
\tau(t)=\frac{t-t_0}{t_{\rm E}}+\delta \tau(t) \qquad ; \qquad \beta(t)=u_0+\delta \beta(t)
\label{eqn:u1u2bis}
\end{equation}
where
\begin{equation}
(\delta \tau(t),\delta \beta(t))=\bpi_{\rm E}\bdv{\Delta p_\odot}=(\bpi_{\rm E}.\bdv{\Delta p_\odot}, \bpi_{\rm E} \times \bdv{\Delta p_\odot)}
\label{paral}
\end{equation}
and $\mathbf{\Delta p_\odot}$ is the apparent position of the Sun
relative to what it would have been assuming rectilinear motion of
the Earth.

The configuration with parallax effects corresponds to Model 2 of
Table~1, 
The resulting diagram showing the north and east
components of $\bpi_{\rm E}$ is presented in
Figure~\ref{fig:par_final}. Taking the parallax effect into account
substantially improves the fit ($\Delta\chi^2 = -52$). The best fit
allowing only for parallax is $\bpi_{\rm E}=(-1.38,0.60)$.  There is a
hard $3\sigma$ lower limit $\pi_{\rm E}>0.6$ and a $3\sigma$ upper
limit $\pi_{\rm E} < 1.9$. If taken at face value, these results would
imply $0.025 < M/M_\odot < 0.075$, i.e., a brown dwarf host with a gas
giant planet.  However, as can be seen from
Figure~\ref{fig:par_final}, these results are
inconsistent with the results from Model 4, which takes
account of both parallax and orbital motion.  This inconsistency
reflects an incorrect assumption in Model 2, namely that the planet is
not moving.

\begin{figure}[h!]
\centering
\includegraphics[width=3in]{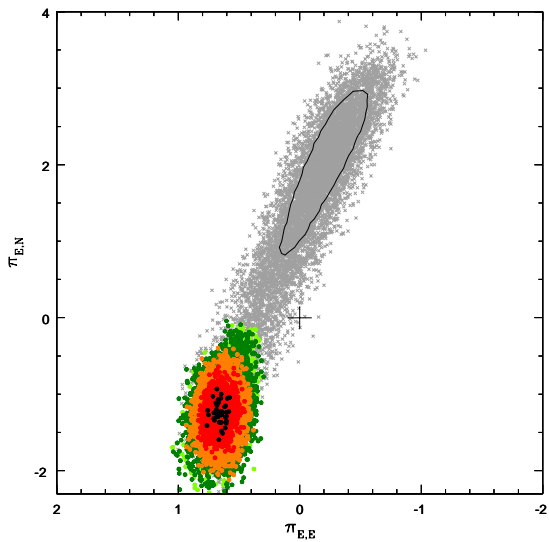}
\caption{\label{fig:par_final}
The $\bpi_{\rm E}$ contours at 1, 2, 3, and $4\,\sigma$ in black, red, orange,
and green, respectively. As a comparison, the gray points show the
approximate $3\sigma$ region of Model 4, i.e., with both parallax and
orbital motion effects, with the $1\sigma$ contour shown in black.
The black cross shows the (0,0) coordinates.
}\end{figure}

\subsection{Orbital motion effects}
\label{orbit}
For the planet orbital motion, we use the formalism of
\cite{dong09}. The lightcurve is capable of constraining at most
two additional orbital parameters that can be interpreted as the
instantaneous velocity components in the plane of the sky. They are
implemented via two new MCMC parameters $ds/dt$ and $\omega$, which
are the uniform expansion rate in binary separation $s$
and the binary rotation rate $\alpha$,
\begin{equation}
s = s_0 + ds/dt\,(t-t_0) \\
\alpha=\alpha_0 + \omega\,(t-t_0).
\end{equation}

These two effects induce variations in the shape and orientation of the
resonant caustic, respectively. To ensure that the resulting orbital
characteristics are physically plausible, we can verify for
any trial solution that the
projected velocity of the planet is not greater than the escape
velocity of the system, $v_{\perp}<v_{esc}$ for a given assumed mass
and distance, where \citep{dong09}
\begin{equation}
 v_{\perp}=\sqrt{(ds/dt)^2+(\omega s)^2}D_l\theta_E
\label{eqn:vperp}
\end{equation}
and
\begin{equation}
 v_{esc}=\sqrt{\frac{2GM}{r}}\leq
v_{esc,\perp}=\sqrt{\frac{2GM}{r_{\perp}}},\qquad
r_{\perp}=s\theta_ED_l.
\label{eqn:vescperp}
\end{equation}

The configuration with only orbital motion corresponds to the Model 3
of Table~1. 
The resulting diagram showing the solution
for the two orbital parameters $\omega$ and $ds/dt$ is presented in
Figure~\ref{fig:rot_final}. Taking the orbital motion of
the planet into account substantially improves the fit ($\Delta\chi^2 =
-67.5$). 
\begin{figure}[h!]
\centering
\includegraphics[width=3in]{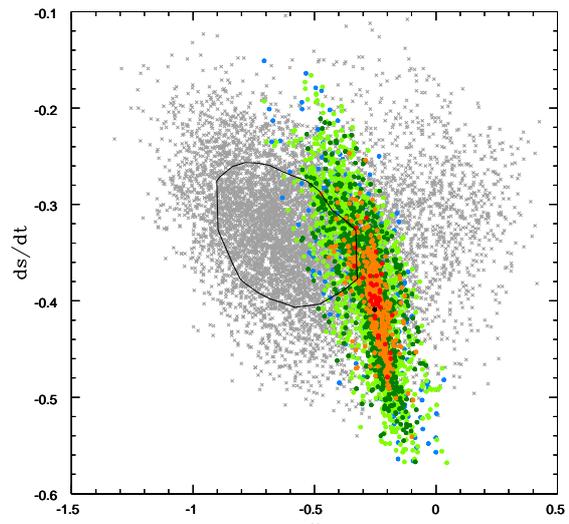}
\caption{\label{fig:rot_final} Orbital parameters of solutions at
1, 2, 3, and $4\,\sigma$ in black, red, orange, and green, respectively. As a
comparison, the gray points show the $3\sigma$ region of Model 4,
i.e., with both parallax and orbital motion effects, with the $1\sigma$
contour shown in black.  }\end{figure}

{\subsection{Combined parallax and orbital motion}
\label{sec:combinedmot}}

In this section we model both parallax and orbital motion effects,
which is called Model 4 in Table~1. 
Taking
these two effects into account results in only a modest improvement in $\chi^2$
compared to the cases for which the effects are considered
individually ($\chi^2_{both} - \chi^2_{orbital} =
-9$). The triangle diagram presented in Figure \ref{fig:triangle}
shows the 2-parameter contours between the four MCMC parameters
$\pi_{\rm{E,N}}$, $\pi_{\rm{E,E}}$, $\omega$ and $ds/dt$ introduced in
Sections \ref{parallax} and \ref{orbit}.  The best fit is
$(\pi_{\rm{E,N}},\pi_{\rm{E,E}})=(2.495,-0.311)$ and
$(\omega,ds/dt)=(-0.738,-0.360)$. This would lead to a host star of
$0.015\,M_\odot$ at a distance $D_l=1.11$ kpc and a $0.21$ Jupiter
mass planet with a projected separation of $0.32$ AU.
 
This small improvement in $\chi^2$ can be explained by a degeneracy
between the north component of $\bpi_{\rm E}$ and the orbital
parameter $\omega$, as shown in Figure \ref{fig:triangle}.  In fact,
the actual degeneracy is between $\pi_{\rm{E},\perp}$ and $\omega$,
where $\pi_{\rm{E},\perp}$ (described by \citealt{gould04}) is the
component of $\bpi_{\rm E}$ that is perpendicular to the instantaneous
direction of the Earth's acceleration, i.e., that of the Sun projected
on the plane of the sky at the peak of the event. This acceleration
direction is $\phi=257.4^\circ$ (north through east). Hence, the
perpendicular direction is $\phi-90^\circ=192.6^\circ$, which is quite
close to the $195.7^\circ$ degeneracy direction in the
$\pi_{\rm{E,N}}$ and $\pi_{\rm{E,E}}$ diagram.
Since $\pi_{\rm{E,\perp}}$ is very close (only $13^\circ$) from north,
$\pi_{\rm{E,N}}$ is a good approximation for it.

Indeed, $\pi_{\rm{E,\parallel}}$ generates an asymmetry in the
lightcurve because, to the extent that the source-lens motion is in
the direction of the Sun-Earth axis, the event rises faster than it
falls (or vice versa). This effect is relatively easy to detect. But
to the extent that the motion is perpendicular to this axis, the Sun's
acceleration induces a parabolic deviation in the trajectory. To
lowest order, this produces exactly the same effect as rotation of the
lens geometry (which is a circular deviation). 
Hence, the degeneracy between $\pi_{\rm{E,\perp}}$ and
$\omega$ can only be broken at higher order. This degeneracy was
discussed in the context of point lenses in \cite{gouldmiralda94},
\cite{smithmaopac03}, and \cite{gould04}. In the point-lens case, the
$\pi_{\rm{E,\perp}}$ degeneracy appears nakedly (because the lens
system is invariant under rotation). In the present case, the
rotational symmetry is broken. In case orbital motion is ignored, it thus
may appear that parallax is measured more easily in binary events, as
originally suggested by \cite{angould01}. But in fact, as shown in the
present case, once the caustic is allowed to ``rotate'' (lowest order
representation of orbital motion), then the $\pi_{\rm{E,\perp}}$
degeneracy is restored.  

\subsection{Limb-darkening implementation}
\label{ld}

Most of the calculations in this paper are done using Stokes' theorem,
which greatly speeds up the computations by reducing a 2-dimensional
integral to one dimension. However, this method implicitly assumes that the
source has uniform surface brightness, whereas real sources are limb
darkened. In the linear approximation, the normalized surface
brightness can be written
\begin{equation}
W(z;\Gamma) = 1-\Gamma\biggl(1-\frac{3}{2}\sqrt{1-z^2}\biggr),
\end{equation}
where $\Gamma$ is the limb-darkening coefficient depending on the
considered wavelength, and $z$ is the position on the source divided by
the source radius. 

We adopt this approach because we expect that the solutions with
and without limb darkening will be nearly identical, except that
the uniform source should appear smaller by approximately a factor
\begin{equation}
{\rho_{\rm uni}\over \rho_{\rm ld}}\simeq
\sqrt{\int dz^2 z^2 W(z:\Gamma)\bigg/\int dz^2 z^2}
= \sqrt{1 - {\Gamma\over 5}}
\label{eqn:ld_ansatz}
\end{equation}
because this ratio preserves the rms radial distribution of light.

To test this conjecture, we approximate the surface as a set of
20 equal-area rings, with the magnification of each ring still computed by
Stokes' method.  The surface brightness of the $i$th ring is
simply $W(z_i)$ where $z_i$ is the middle of the ring.  The
limb-darkening coefficients for the unfiltered data have been
determined by interpolation, from $V$, $R$, $I$ and $H$ limb-darkening
coefficients.  We find from the CMD that the source star has
$(V-I)_0=0.82$, so roughly a G7 dwarf or slightly cooler. We adopt a
temperature of $T=5500$ K. We thus obtain the following limb-darkening
parameters $(u_V,u_R,u_I,u_H)=(0.7117,0.6353,0.5507,0.3659)$, where
$u=3\Gamma/(\Gamma+2)$ \citep{afon00}. Then
$(\Gamma_V,\Gamma_R,\Gamma_I,\Gamma_H)=(0.6220,0.5373,0.4497,0.2778)$.
For a given observatory/filter (or possibly unfiltered), 
we then compare 
$(R_{observed}-I_{CTIO})$ to $(V_{CTIO}-I_{CTIO})$, considering that
$I_{CTIO}=0.07V+0.93I$ and that approximately $V=2R-I$ and deduce
empirical expression for the corresponding $\Gamma$ coefficients.
The $\Gamma$ coefficients for all the observatories then
become
$(\Gamma_{MOA},\Gamma_{SAAO},\Gamma_{FCO},\Gamma_{AO},\Gamma_{Danish},
\Gamma_{Bronberg},\Gamma_{Wise})=(0.493,0.45,0.52,0.51,0.45,0.53,0.49)$.
Substituting, a mean $\Gamma\sim 0.47$ into Eq.~(\ref{eqn:ld_ansatz}),
we expect $\rho$ to be $\sim 5\%$ larger when limb-darkening
is included.

{\subsection{Results summary}
\label{results1}}
We summarize the best-fit results for the five different models presented 
in Section~\ref{sec:model} in Table~1. 
The five models are Model 1: Finite-source binary-lens model with neither parallax nor orbital motion effects; Model 2: Finite-source binary-lens model with parallax effects only; Model 3: Finite-source binary-lens model with orbital motion effects only; Model 4: Finite-source binary-lens model with both parallax and orbital motion effects; and Model 5: Finite-source binary-lens model with both parallax and orbital motion effects and limb-darkening.

Note in particular that Models 4 and 5 agree within $\sim 1\,\sigma$
for all parameters, except that $\rho$ is $\sim 7\%$ greater in the limb-darkened
case (Model 5).

{\section{Bayesian analysis}
\label{sec:bayesian}}

The Markov Chain used to find the solutions illustrated
in Figure \ref{fig:triangle} is constructed (as usual)
by taking trial steps that are uniform in the MCMC variables,
including $t_0$, $u_0$, and $t_{\rm E}$.  This amounts to
assuming a uniform prior in each of these variables. In the
case of the three variables $t_0$, $u_0$, and $t_{\rm E}$, the solution is
extremely well constrained, so it makes hardly any difference
which prior is assumed. Whenever this is the case, Bayesian and
frequentist orientations lead to essentially the same results.
However, as shown in Figure \ref{fig:triangle}, $\vec\pi_{\rm E}$ is quite
poorly constrained: at the $2\sigma$ level, the magnitude of $\pi_{\rm E}$
varies by more than an order of magnitude.  Since the lens
distance is related to the microlens parallax by
$D_l= {\rm AU}/(\theta_{\rm E}\pi_{\rm E}+\pi_S)$, where 
$\pi_S={\rm AU}/D_s$,
this amounts to giving equal prior weight to a tiny range of distances
nearby and a huge range of distances far away.  But the actual weighting
should have the reverse sign, primarily because a fixed distance range
corresponds to far more volume at large than small distances.  In fact,
a Galactic model should be used to predict the a priori expected
rate of microlensing events, which depends not only on the correct volume
element but also on the density and velocity distributions of the 
lens and the source as well.

Similarly, a Keplerian orbit can be equally well
characterized by specifying the seven standard Kepler parameters
or six phase-space coordinates at a given instant of time, plus
the host mass. The latter parametrization is more convenient from a
microlensing perspective because microlensing most robustly measures
the two in-sky-plane Cartesian spatial coordinates
($s\cos\alpha$ and $s\sin\alpha$) and the two in-plane Cartesian velocity
coordinates ($ds/dt$ and $s\omega$), while the mass is directly
given by microlens variables $M=\theta_{\rm E}/\kappa\pi_{\rm E}$.
However, the former (Kepler) variables have simple well-established
priors. By stepping equally in microlens parameters, one is
effectively assuming uniform priors in these variables, whereas
one should establish the priors according to the Kepler parameters.

In principle, one would simultaneously incorporate both sets
of priors (Galactic and Kepler), and we do ultimately adopt
this approach. However, it is instructive to first apply
them separately to determine whether these two sets of priors
are basically compatible or are relatively inconsistent.

\onecolumn{
\begin{figure}[h!]
\centering
\includegraphics[width=5in]{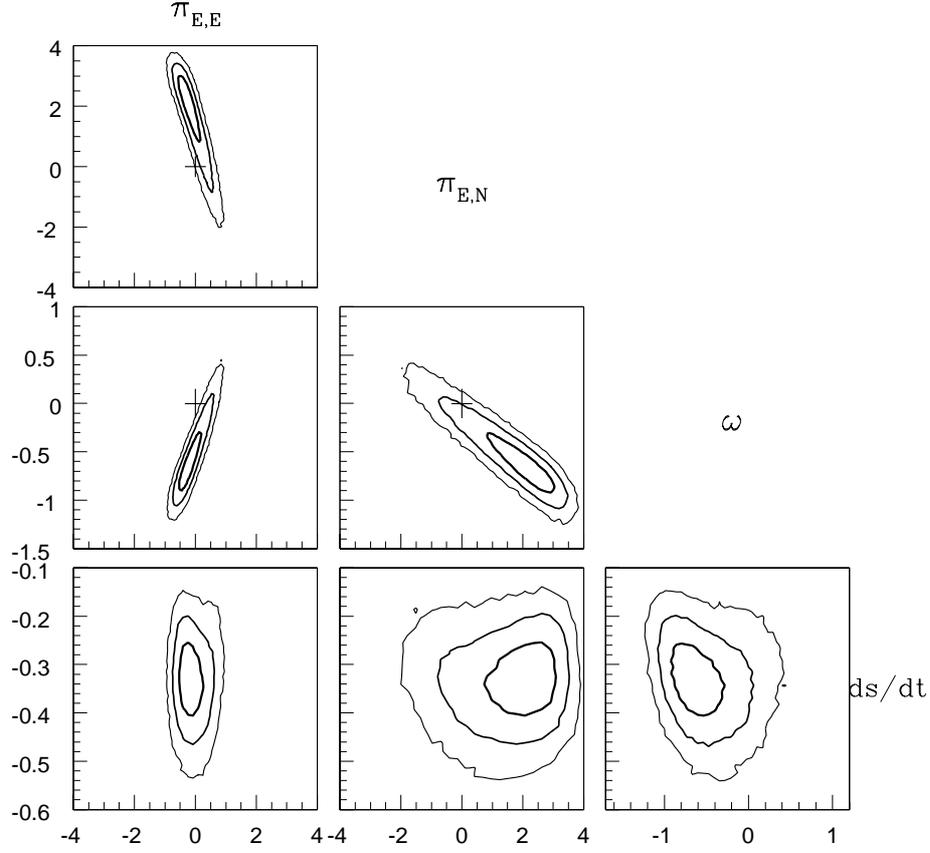}
\caption{\label{fig:triangle} Parallax and orbital motion parameters
of solutions contours at 1, 2, 3, and $4\,\sigma$. The black crosses show the (0,0) coordinates. }\end{figure}

\begin{table}[h]
\begin{center}
\begin{tabular}{l r r r r r r r r r r r}

\multicolumn{12}{c}{Table 1. Fit parameters for finite-source binary-lens models} \\
\hline
\hline
Model & $t_0$ & $u_0$ & $t_E$ & $s$ & $q$ & $\alpha$ & $\rho$ & $\pi_{E,N}$ & $\pi_{E,E}$ & $\omega$ & $ds/dt$\\ 
\multicolumn{1}{l}{$\chi^2$} & \multicolumn{11}{c}{Error bars}\\
\hline
Model 1 & 5042.34 & 0.0683 & 48.7 & 0.9152 & 0.01073 & 4.3074 & 0.00149 & - & - & - & - \\
1100 & 0.01 & 0.0005 & 0.4 & 0.0002 & 0.00015 & 0.0025 & 0.00002 & - & - & - & - \\
\hline
Model 2 & 5042.38 & 0.0770 & 43.9 & 0.9137 & 0.01230 & 4.3063 & 0.00174 & -1.38 & 0.60 & - & - \\
1048 & 0.02 & 0.0015 & 0.5 & 0.0004 & 0.00030 & 0.0030 & 0.00005 & 0.25 & 0.07 & - & - \\
\hline
Model 3 & 5042.32 & 0.0902 & 38.4 & 0.9137 & 0.0135 & 4.302 & 0.00197 & - & - & -0.252 & -0.409\\
1032.5 & 0.02 & 0.002 & 0.6 & 0.0003 & 0.0002 & 0.002 & 0.00005 & - & - & 0.1 & 0.04 \\
\hline
Model 4 & 5042.366 & 0.0890 & 40.1 & 0.9134 & 0.0135 & 4.3095 & 0.00195 & 2.5 & -0.31 & -0.74 & -0.36 \\
1024.5 & 0.015 & 0.0010 & 0.5 & 0.0002 & 0.0002 & 0.0025 & 0.00003 & 1 & 0.3 & 0.2 & 0.05 \\
\hline
Model 5 & 5042.36 & 0.0881 & 40.0 & 0.9136 & 0.0132 & 4.3099 & 0.00202 & 1.7 & -0.15 & -0.51 & -0.37 \\
1029.2 & 0.02 & 0.0010 & 0.5 & 0.0003 & 0.0002 & 0.0025 & 0.00003 & 1 & 0.5 & 0.3 & 0.05 \\
\hline
\end{tabular}
\end{center}
\end{table}
}

\twocolumn{

Formally, we can evaluate the
posterior distribution $f(X\mid D)$, including both {\it prior}
expectations from (Galactic and/or Keplerian) models
and {\it posterior} observational data 
using Bayes' Theorem~:
\begin{equation}
f(X\mid D)=\frac{f(D\mid X)f(X)}{f(D)}.
\label{eqn:bayes}
\end{equation}
Here $f(D\mid X)$ is the likelihood function over the data
$D$ for a given model $X$, $f(X)$ is the prior distribution
containing all \textit{ex ante} information about the parameters $X$
available before observing the data, and $f(D)=\int_X f(D\mid X)f(X)dX$.
In the present context, this standard Bayes formula is interpreted
as follows: the density of links on the MCMC chain directly gives
 $f(D\mid X)$, while $f(X)$ encapsulates the parameter priors, including
both the underlying rate of events in a ``natural physical coordinate system''
in which these priors assume a simple form
and the Jacobian of the transformation from this ``physical'' system
to the ``natural microlensing parameters'' that are directly modeled
in the lightcurve analysis.

It is not obvious, but we find below that the coordinate
transformations for Galactic and Kepler models actually
factor, so we can consider them independently.

\subsection{Galactic model}
Applying the generic rate formula $\Gamma = n\sigma v$ to
microlensing rates as a function of the independent physical variables
$(M,D_l,\bmu)$, yields
\begin{equation}
f_{\rm Gal}(X) \propto
\frac{d^4\Gamma}{d D_L\,dM\,d^2\mu}=\nu(x,y,z)(2 R_{\rm E})v_\rel f(\bmu) g(M),
\label{eqn:fxgal}
\end{equation}
where the spatial positions $(x,y,z)$, the physical Einstein radius
$R_{\rm E}$, and the lens velocity relative to the observer-source
line of sight $v_\rel$ are all regarded as dependent variables
of the four variables shown on the l.h.s., plus the two angular
coordinates.  Here $\nu(x,y,z)$ is the local density of lenses,
$g(M)$ is the mass function [we will eventually adopt $g(M)\propto M^{-1}$],
and $f(\bmu)$ is the two-dimensional probability function for a
given source-lens relative proper motion, $\bmu$.  Since
$v_\rel = \mu D_l$ and $R_{\rm E} = D_l\theta_{\rm E}$, this can be
rewritten in terms of microlensing variables,
$$
{d^4\Gamma\over d t_{\rm E}\,d\theta_{\rm E}\,d^2\pi_{\rm E}} =
{d^4\Gamma\over d D_L\,dM\,d^2\mu}\,
\times{\mu\over\pi_{\rm E}}
\bigg|{\partial(D_L,M,\mu)\over\partial(t_{\rm E},\theta_{\rm E},\pi_{\rm E})}\bigg|
$$
$$
= 2 D_l^2\theta_{\rm E}\mu\nu(x,y,z) f(\bmu) g(M)\times {2\over {\rm AU}} D_l^2
{M\,\pi_\rel\,\mu^2\over t_{\rm E}\theta_{\rm E}\pi_{\rm E}^2},
$$
where $M=\theta_{\rm E}/\kappa\pi_{\rm E}$, $D_l={\rm AU}/(\pi_\rel + \pi_s)$,
$\pi_\rel=\theta_{\rm E}\pi_{\rm E}$, and $\mu=\theta_{\rm E}/t_{\rm E}$
are now regarded as dependent variables. We note that 
$$ {\partial(D_L,M,\mu)\over\partial(t_{\rm E},\theta_{\rm E},\pi_{\rm
E})}={\partial(\pi_\rel,M,\mu)\over\partial(t_{\rm E},\theta_{\rm
E},\pi_{\rm E})}{dD_L \over d\pi_\rel}={2\pi_\rel\,M\, \mu \over
t_E\, \theta_{\rm E}\, \pi_{\rm E}} {D_L^2 \over {\rm AU}},
$$
where the last evaluation follows from the general theorem:
$$
y_i = \prod_j x_j^{\alpha_{ij}}\Longrightarrow 
{\partial (y_i)\over \partial (x_j)} = 
{\partial (\ln y_i)\over \partial (\ln x_j)}{\prod_i y_i\over \prod_j x_j} = 
|\alpha|{\prod_i y_i\over \prod_j x_j}.
$$

Finally, Eq.~(\ref{eqn:fxgal}) reduces to
\begin{equation}
\label{eqn:gamma}
{d^4\Gamma\over d t_{\rm E}\,d\theta_{\rm E}\,d^2\pi_{\rm E}} =
{4\over \rm AU} \nu(x,y,z)f(\bmu) [g(M)M]
{D_l^4\mu^4\over\pi_{\rm E}}.
\end{equation}

The variables on the l.h.s. of Eq.~(\ref{eqn:gamma}) are
essentially the Markov chain variables in the microlensing fit
procedure. \footnote{In fact, $\rho$ is used in place of $\theta_{\rm E}$, but
this makes no difference, since $\theta_{\rm E}\propto \rho$).} 
The distribution of MCMC links applied to
the data can be thought of as the posterior probability distribution 
of the Markov-chain
variables {\it under the assumption that the prior probability
distribution in these variables is uniform.}  In our case, the prior
distribution is not uniform, but is instead given by the r.h.s. of
Eq.~(\ref{eqn:gamma}).  We therefore must weight the output of the
MCMC by this quantity, which is the specific evaluation of
$f(X)$ in Eqs.~(\ref{eqn:bayes}) and (\ref{eqn:fxgal}).

As mentioned above, we adopt $g(M)\propto M^{-1}$, so the term in
square brackets disappears.  We evaluate $\nu(x,y,z)$ and
$f(\bmu)$ as follows.

\subsubsection{Lens-source relative proper motion distribution $f(\bmu)$}

To compute the relative proper motion probability, 
we assume that the velocity distributions of the lenses and sources
are Gaussian
$f(v_y,v_z)=f(v_y)f(v_z)$ where
\begin{equation}
\label{eqn:gauss}
f(\mu_y)=f(v_y)\frac{dv_y}{d\mu_y}=D_L\frac{1}{\sqrt{2\pi\sigma_y^2}}\rm{exp}\bigg[-\frac{(v_y-\tilde{v}_y)^2}{2\tilde{\sigma_y}^2}\bigg]
\end{equation}
and a similar distribution for $f(\mu_z)$. 
Here $v_y$ and $v_z$ are components of the projected velocity $\vec{v}$ derived from the MCMC fit, which is expressed by $\vec{v}=\vec{\mu}D_l$, where
\begin{equation}
 \vec{\mu}=\frac{\vec{\pi_E}}{\pi_E}\frac{\theta_E}{t_E}.
\end{equation}
The expected projected velocity which appears in Eq.\ref{eqn:gauss} is defined as
\begin{equation}
\bdv{\tilde{v}}=\bdv{v_l}-\Bigg[\bdv{v_s}\frac{D_{l}}{D_{s}}+\bdv{v_o}\frac{D_{ls}}{D_{s}}\Bigg]
\end{equation}
where $D_l$, $D_s$ are respectively the lens and source distances from
the observer and $D_{ls}$ the lens-source distance. The velocity is
expressed in the $(x,y,z)$ coordinate system, centered on the center
of the Galaxy, where $x$ and $z$ axes point to the Earth and the North
Galactic pole, respectively. As given in \cite{hangould95}, we adopt
$\vec{v}_{z,disk}=\vec{v}_{z,bulge}=0$ and
$\sigma_{z,disk}=20\,\rm{km.s}^{-1}$,
$\sigma_{z,bulge}=100\,\rm{km.s}^{-1}$ for the $z$ component of the
velocity. For the $y$ direction,
$\vec{v}_{y,disk}=220\,\rm{km.s}^{-1}$, $\vec{v}_{y,bulge}=0$ and
$\sigma_{y,disk}=30\,\rm{km.s}^{-1}$,
$\sigma_{y,bulge}=100\,\rm{km.s}^{-1}$ depending on whether the lens is
situated in the disk or in the bulge. We also consider the asymmetric
drift of the disk stars by subtracting $10\,\rm{km.s}^{-1}$ from
$\vec{v}_{y,disk}$. The celestial north and east velocities of the
Earth seen by the Sun at the time of the event are
$\vec{v}_E=(v_{E,E},v_{E,N})=(+22.95,-3.60)\,\rm{km.s}^{-1}$. In the
Galactic frame, the galactic north and east components of the Earth
velocity become
\begin{equation}
\label{eqn:earthvel}
v_{\rm E,North\,Gal}=v_{E,N}\cos{59.7^\circ}-v_{E,E} \sin{59.7^\circ},
\end{equation}
\begin{equation}
v_{\rm E,East\,Gal} = v_{E,N}\sin{59.7^\circ}+v_{E,E} \cos{59.7^\circ}.
\end{equation}
The velocity of the Sun in the Galactic frame is
$\vec{v}_\odot=(7,12)\,\rm{km.s}^{-1} + (0,v_{\rm{circ}})$, where
$v_{\rm{circ}}=220\,\rm{km.s}^{-1}$, from which 
we deduce the velocity $v_o$ of the observer in the Galactic frame
by adding the Earth velocity from Eq.~(\ref{eqn:earthvel}).

\subsubsection{Density distribution $\nu(x,y,z)$}
The density distribution, $\nu(x,y,z)$, is given at the lens coordinates
(x,y,z) in the Galactic frame.  For this distribution, 
we adopt the model of \citet{hangould03}, which
is based primarily on star counts, and, without any adjustment,
reproduces the microlensing optical depth measured toward
Baade's window. The density models are given in Table~\ref{tab:density}. The disk parameters are
$H=2.75\,$kpc, $h_1=156\,$pc, $h_2=439\,$pc, and $\beta=0.381$, 
where $R\equiv(x^2+y^2)^{1/2}$.  
For the barred (anisotropic) bulge model,
$r_s=\big([(x'/x_0)^2+(y'/y_0)^2]^2+(z'/z_0)^4 \big) ^{1/4}$. Here the
coordinates $(x',y',z')$ have their center at the Galactic center, the
longest axis is the $x'$, which is rotated $20^\circ$ from the Sun-GC axis
toward positive longitude, and the shortest axis is the $z'$ axis. The
values of the scale lengths are $x_0=1.58$ kpc, $y_0=0.62$ kpc and
$z_0=0.43$ kpc respectively. For the bulge, \citet{hangould03} normalize
the ``G2'' K-band integrated-light-based bar model of 
\citet{dwek95} using star counts toward 
Baade's window from \citet{holtzman98} and \citet{zoccali00}.
For the disk, they incorporate the model of \citet{zheng01}, which
is a fit to star counts.

In the calculation, we sum the probabilities of disk and
bulge locations for the lens. We set the limits of the disk range to
be $[0,7]$ kpc from us and $[5,11]$ kpc for the bulge range. We also
apply the bulge density distribution to the source, in the $[6.5,11]$
kpc range.
Rigorously, because we already know the dereddened flux of the source, we should have derived a distribution of sources from the luminosity distribution  of bulge stars combined with their distance. However, as we do not know  the precise distribution of bulge luminosities at fixed color, we only consider the density distribution of sources as a function of their position in the bulge only. Because the stellar density drops off very rapidly from the peak,
the source is effectively localized as being close to the Galactocentric
distance.

\subsection{Orbital motion model}
In addition to the Galactic model, we build a Keplerian model to put
priors on the orbital motion of the planet.  To extract the orbital
parameters from the microlensing parameters, we refer to the appendix
of \cite{dong09}.  Given that from the light curve of the
event we have access to the instantaneous projected velocity and
position of the planet for only a short time, we consider a
circular orbit to model the planet motion.  The distortions of the
light curve are modeled by $\omega$ and $ds/dt$, which then specify
the variations in orientation and shape of the resonant caustic,
respectively. These quantities are
defined in Section~\ref{orbit}.  Since $r_{\perp}=D_l\theta_E d$ 
is the projected star-planet separation, we evaluate the
instantaneous planet velocity in the sky plane, with
$r_{\perp}\gamma_{\perp}=r_{\perp}\omega$ the velocity perpendicular
to the planet-star axis and $r_{\perp}\gamma_{\parallel}=r_{\perp}(ds/dt)/s$ 
the velocity parallel to this axis. We define the $\hat{i},\hat{j},\hat{k}$
directions as the instantaneous star-planet axis on the sky plane, the
direction into the sky, and $\hat{k}=\hat{i}\times\hat{j}$. In this
frame, the planet is moving among two directions, defined by the
angles $\theta$ and $\phi$, which are effectively a (complement to a) polar 
angle and an azimuthal angle, 
respectively. Specifically, $\phi$ is the angle between the
star-planet-observer ($r_{\perp}=a\sin\phi$), and $\theta$ characterizes
the motion in the direction of the velocity along $\hat{k}$. Then the
instantaneous velocity of the planet is
\begin{equation}
 v=\sqrt{\frac{GM}{a}}[\cos\theta\hat{k}+\sin\theta(\cos\phi\hat{i}-\sin\phi\hat{j})]
\end{equation}
where $a$ is the semimajor axis.  Thus we obtain
$\gamma_{\perp}=\sqrt{\frac{GM}{a^3}}\frac{\cos\theta}{\sin\phi}$ and
$\gamma_{\parallel}=\sqrt{\frac{GM}{a^3}}\sin\theta\cot\phi$.  The
Jacobian expression to transform from
$P(s,\gamma_{\perp},\gamma_{\parallel})$ to $P(a,\phi,\theta)$
is 


\begin{equation}
\label{eqn:jac}
J={\partial(a,\phi,\theta)\over
\partial(s,\gamma_{\perp},\gamma_{\parallel})}
={a^3\over GM}
\tan^2\phi \big(\frac{1}{2}-\sin^2\theta\tan^2\phi \big)^{-1}R_{\rm E}
\end{equation}

As explained in \cite{dong09}, for one set of microlensing parameters,
there are two degenerate solutions in physical space. In the
orbital model, we consider the two
solutions to constrain the light curve fit, each with its own separate probability.

From the definition of the two angles, the transformation of the
polar system $(a,\pi/2-\theta,\phi)$ contains the quantity
$\sin\theta$ and so the Jacobian includes the factor $\cos\theta$ from
$d(\sin\theta) d\phi= d\theta d\phi \cos\theta$. Moreover, we adopt
a flat distribution on $\ln(a)$, implying the factor $1/a$ in the
Jacobian expression.  Then,
\begin{equation}
J=\frac{\partial(\ln(a),\phi,\sin\theta)}
{\partial(s,\gamma_{\perp},\gamma_{\parallel})}
=\frac{r^2_{\perp}}{GM}\frac{\cos\theta}{\cos^2\phi}
\big(\frac{1}{2}-\sin^2\theta\tan^2\phi \big)^{-1} R_{\rm E}
\label{eqn:jacprime}
\end{equation}
Note that the terms $\sin\theta$ and $\cos\theta$ in the denominators
of Eq.~(\ref{eqn:jacprime}) correct an error in \cite{dong09}.

\subsection{Constraints from VLT}

As foreshadowed in Section~\ref{sec:VLT}, the VLT NACO flux
measurement places upper limits on the flux from the lens, hence on its mass (assuming it is not a white dwarf).
However, we begin by assuming that the excess light is caused by the
lens.  We do so for two reasons.  First, this is actually the
most precise way to enforce an upper limit on the lens flux.
Second, it is of some interest to see what mass range is
``picked out'' by this measurement, assuming the excess flux
is due to the lens.  

The first point to note is that, if the lens contributes any
significant flux, then it lies behind most or all of the dust
seen toward the source.  For example, if the lens mass is just
$M=0.15\,M_\odot$ (which would make it quite dim, $M_H>8$), then
it would lie at distance 
$D_L = {\rm AU}/(\theta_{\rm E}^2/\kappa M + {\rm AU}/D_S)
= 4.9\,$kpc, where we have adopted the central values
$\theta_{\rm E} = 0.31\,$mas and $D_S=8.7\,$kpc for this
exercise.  More massive lenses would be farther.

Next we estimate $A_H=0.4$ from the measured clump color $(V-I)_{\rm cl}=2.10$,
assuming an intrinsic color of the red giant clump
of $(V-I)_{0,\rm cl}=1.08$ \citep{bensby10} 
and adopting for this line of sight $A_H/E(V-I) = 0.40$.

Finally, for the relation between $M$ and $M_H$, we consult the
library of empirically-calibrated isochrones of \cite{an07}.  We adopt
the oldest isochrones available (4 Gyr), since there is virtually no
evolution after this age for the mass range
that will prove to be of interest $M<0.7\,M_\odot$.  Moreover,
in this mass range, the isochrones hardly depend on metallicity
within the range explored ($-0.3<{\rm [Fe/H]}<+0.2$).  

\onecolumn{
\begin{table}[h]
\begin{center}
\begin{tabular}{l l l}
\multicolumn{3}{c}{Table 2. Density distribution for the bulge and disk models} \\
\hline
\hline
\multicolumn{1}{c}{Location} & \multicolumn{1}{c}{Model} & \multicolumn{1}{c}{Distribution (in $M_\odot\,{\rm pc}^{-3}$)}\\
\hline
Bulge & Dwek & $\nu(r_s) = 1.23 \exp (-0.5r^2_s)$\\
Disk  & Zheng & $\nu(R,\, z)= 1.07\,\exp(-R/H)[(1-\beta)(-|z|/h_1) + \beta \exp(-z/h_2)]$\\
\hline
\end{tabular}
\end{center}
\label{tab:density}
\end{table}

\begin{figure}[h!]
\centering
\includegraphics[width=6in]{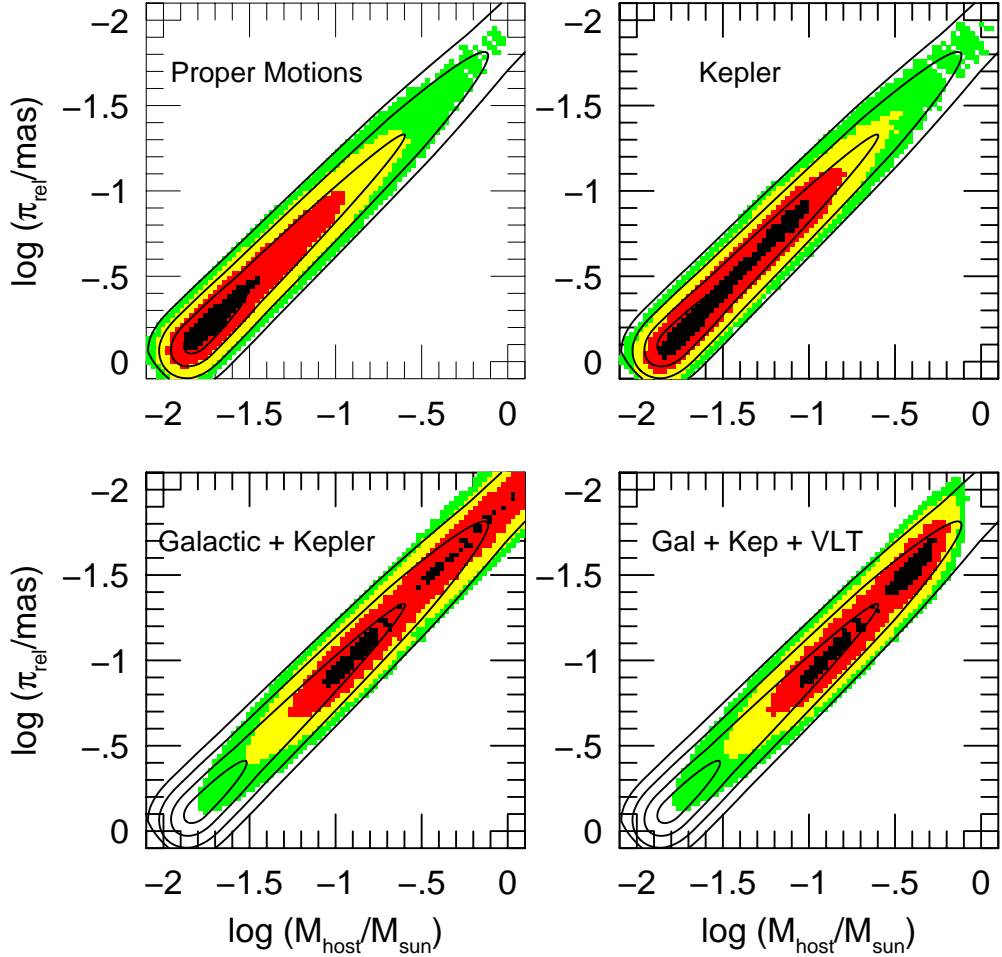}
\caption{\label{fig:mpirel} Bayesian analysis results.
Each panel shows host mass $M$ versus lens-source relative
parallax $\pi_{\rm rel}$, with 1, 2, 3, and $4\,\sigma$ contours
under two different conditions.  The solid black contours
are derived from the light curve alone, without any priors.
The colored symbols show contour levels after applying various
priors, respectively Galactic proper motion only, Kepler only,
full Galactic and Kepler priors, and full Galactic and Kepler priors,
plus VLT imaging constraints.
The proper-motion and Kepler priors are fully consistent with
the light curve, but there is strong tension between between
the distance-related priors and the lightcurve, with the former
favoring high masses and small lens-source separations.
The highest part of this disputed mass range, $M>0.7\,M_\odot$,
is essentially ruled out by the VLT imaging constraint (lower right).}
\end{figure}
}

\twocolumn{

For each mass and distance considered below, we then calculate
$H_L = M_H + A_H + 5\log(D_L/10\,{\rm pc})$ and combine the
corresponding flux with $H_S = 18.35$ to obtain $H_{\rm pred}$.
We then calculate a likelihood factor 
$L_H=\exp[-(H_{\rm pred} - H_{\rm obs})^2/2\sigma_H^2]$,
where $H_{\rm obs}=18.25$ and $\sigma_H=0.07$, as discussed
in Section~\ref{sec:VLT}.

For fiducial values $D_S=8.7\,$kpc and $\theta_{\rm E}=0.31\,$mas,
this likelihood peaks at $M=0.42\,M_\odot$, but it does so very
gently. The suppression factor is just $L_H\sim 0.7$ at
$M=0.21\,M_\odot$ and $M=0.52\,M_\odot$.  At lower masses, even if there
were zero flux, the suppression would never get lower than $L_H=0.36$,
simply because the excess-flux measurement is consistent with zero
at $1.4\,\sigma$.  But at higher mass, the expected flux quickly
becomes inconsistent.  For example, $L_H(0.65\,M_\odot)= 0.07$.

Hence, by treating the flux measurement as an excess-flux
``detection'', we impose the ``upper limit'' on mass in a graceful
manner.  Moreover, as regards the upper limit, this approach
remains valid when we relax the assumption that the excess flux
is solely due to the lens.  That is, even if there are other
contributors, the likelihood of a given high-mass lens being compatible
with the flux measurement can only go down.

However, the same reasoning does not apply at the low-mass limit.  For
example, if the excess flux came from a source companion or
an ambient star, then a brown-dwarf lens would be fully compatible
with the flux measurement.  Nevertheless, this is quite a minor
effect because, in any event, the suppression factor would not
fall below 0.36.  To account for other potential sources of
light, we impose a minimum suppression factor $L_{H,{\rm min}}=0.5$ at the
low-mass end.

\subsection{Combining Galactic and Kepler priors and adding VLT constraints}

In this section, we impose the priors from the Galactic and Kepler models
and add the constraints from the VLT flux measurement.
We defer the VLT constraints to the end because they
do not apply to the special case of white-dwarf lenses.

We begin by examining the role of the various priors separately
to determine the level of ``tension'' between these and the
$\chi^2$ derived from the light curve alone.  We do so because
each prior involves different physical assumptions, and tension
with the light curve may reveal shortcomings in these assumptions.

The Kepler priors involve two assumptions, first that the
planetary system is viewed at a random orientation (which is almost
certainly correct) and second that the orbit is circular
(which is almost certainly not correct).  We will argue further 
below that the assumption of circular orbits has a modest impact.
In any event, we want to implement the Kepler priors by themselves.

The Galactic priors really involve two sets of assumptions.
The more sweeping assumption is that planetary systems are distributed
with the same physical-location distribution and host-mass distribution
as are stars in the Galaxy.  We really have no idea whether this
assumption is true or not.  For example, it could be that bulge
stars do not host planets.  The assumptions about host mass
and physical location are linked extremely strongly in a mathematical
sense (even if they prove to be unrelated physically) because
$\theta_{\rm E}$ is well-measured, and 
$\theta_{\rm E}^2 = \kappa M\,\pi_{\rm rel}$.
Thus, we must be cautious about this entire set of
assumptions.

However, the Galactic priors also contain another factor $f(\bmu)$,
in which we can have greater a priori confidence.  This prior basically
assumes that planetary systems at a given distance (regardless of
how common they are at that distance) will have similar kinematics
to the general stellar population at the same distance.  The scenarios
in which this assumption would be strongly violated, while not impossible,
are fairly extreme.

Therefore we begin by imposing proper-motion-only and Kepler-only
priors in the top two panels of Figure~\ref{fig:mpirel}, 
which
plots host mass $M$ versus lens-source relative parallax $\pi_{\rm rel}$.
We choose to plot $\pi_{\rm rel}$ rather than $D_L$ because
it is given directly by microlensing parameters 
$\pi_{\rm rel} = \pi_{\rm E}\theta_{\rm E}$.  The 1, 2, 3, and 4 $\sigma$ 
contours from the $\chi^2$ based on the light curve only 
are shown in black.  Each of
these priors is consistent with the light curve at the $1\,\sigma$
level, so we combine them and find that they still
display good consistency.  In  the lower left panel, we combine
the full Galactic and Kepler priors. These tend to favor much
heavier, more distant lenses, which are strongly disfavored by the
lightcurve, primarily because of the factor
$D_l^4/\pi_{\rm rel}$ in Eq.~(\ref{eqn:gamma}). Indeed masses
$M>0.7\,M_\odot$ will be effectively ruled out by high-resolution 
VLT imaging, further below.

When combining Galactic and Kepler priors, we simply weight the
output of the MCMC by the product of the factors corresponding to
each.  This is appropriate because,
while the $6\times 6$ matrix, transforming 
the full set of microlensing parameters
$(s,\gamma_\perp,\gamma_\parallel,t_E,\theta_E,\pi_E)$ to the
full set of physical parameters $(a,\phi,\theta,M,D_L,\mu)$, 
is {\it not} block diagonal, the Jacobian nevertheless factors as
$$
\frac{\partial(a,\phi,\theta,M,D_L,\mu)}
{\partial(s,\gamma_{\perp},\gamma_{\parallel},t_E,\theta_E,\pi_E)}
=
\frac{\partial(a,\phi,\theta)}
{\partial(s,\gamma_{\perp},\gamma_{\parallel})}
\times
\frac{\partial(M,D_L,\mu)}{\partial(t_E,\theta_E,\pi_E)}.
$$
Hence, the full weight, $f(X)$ in Eq.~(\ref{eqn:bayes})
is simply the product of the two found separately for the
Galactic and orbital priors. 

\begin{figure}[h!]
\centering
\includegraphics[angle=0,width=4in]{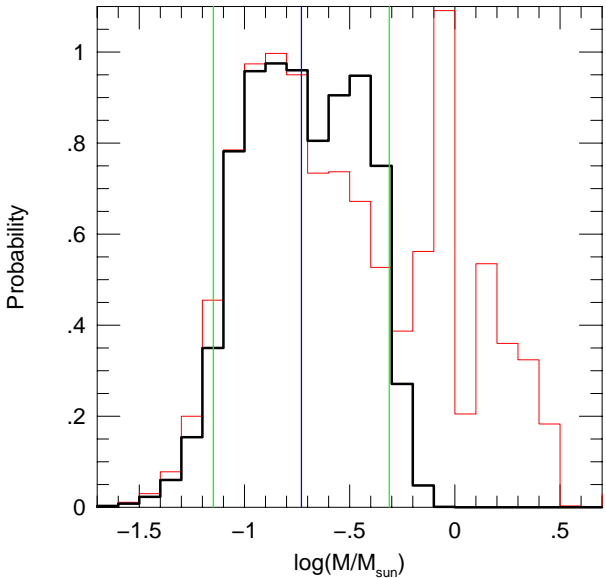}
\caption{\label{fig:mass-alone}
Probability as a function of host mass after applying the
Galactic and Kepler priors ({\it red}) and then adding the
constraints from VLT observations ({\it black}).}
\end{figure}

Figure~\ref{fig:mass-alone} shows the host-mass probability 
distribution before ({\it red}) and after ({\it black}) applying
the constraint from VLT imaging to the previous analysis incorporating
both Galactic and Kepler priors. The 90\% confidence interval is marked. The high mass solutions toward the right are strongly disfavored by the lightcurve (see Figure~\ref{fig:mpirel}), but the Galactic prior for them is so strong that they have substantial posterior probability. However, these solutions are heavily suppressed by the VLT flux limits. The hsot is most likely to be an M dwarf. 
The lower right panel of Figure~\ref{fig:mpirel} shows the 2-dimensional
$(M,\pi_{\rm rel})$ probability distribution for direct comparison
with the results from applying various combinations of priors.

\subsection{Bayesian results for physical parameters}

Table~3 shows the median estimates and 90\% confidence intervals
for six physical parameters (plus one physical diagnostic) as more
priors and constraints are applied.  The bottom row, which includes
full Galactic and Kepler priors, plus constraints from VLT photometry
shows our adopted results.  The six physical parameters are the
host mass $M$, the planet mass  $m_p$, the distance of the system $D_L$,
the period $P$, the semi-major axis $a$, and the orbital inclination $i$.  The
last three assume a circular orbit. For rows 2 and 4
(which do not apply Kepler constraints), the values shown 
for $(P,a,i)$ summarize
the results restricted to links in the chain that are consistent with
a circular orbit, while the first four columns summarize all links in
the chain.  The key results are
\begin{equation}
\label{eqn:massrange}
0.071\,M_\odot < M_{\rm host} < 0.49\,M_\odot\qquad (90\% \ \rm confidence)
\end{equation} 
and corresponding to this, $m_p = q M$, where $q=0.00132\pm 0.00002$, i.e.,
\begin{equation}
\label{eqn:massrange2}
1.0\,M_{\rm Jup} < m_p < 6.7\,M_{\rm Jup} \qquad (90\% \ \rm confidence),
\end{equation} 
\begin{equation}
\label{eqn:perrange}
3.8\,{\rm yr} < P < 7.6\,{\rm yr}\qquad (90\% \ \rm confidence)
\end{equation} 
\begin{equation}
\label{eqn:semirange}
1.1\,{\rm AU} < a < 2.7\,{\rm AU}\qquad (90\% \ \rm confidence)
\end{equation} 
with the medians at $M=0.19\,M_\odot$, $m_p=2.6\,M_{\rm Jup}$, 
$P=5.4\,$yr, $a=1.8\,$AU.  That is, the host is an M dwarf with 
a super-Jovian massive planetary companion. For completeness,
we note that in obtaining these results, we have implicitly
assumed that the probability of a star having a planet with
a given planet-star mass ratio $q$ and semi-major axis $a$ is
independent of the host mass and distance.

\subsection{White dwarf host?}

When we applied the VLT flux constraint, we noticed that it would not
apply to white-dwarf hosts.  Is such a host otherwise permitted?
In principle, the answer is ``yes'', but as we now show, it is
rather unlikely.  The WD mass function peaks at about $M\sim 0.6\,M_\odot$,
which corresponds to an $M_{\rm prog}\sim 2\,M_\odot$ progenitor.
If the progenitor had a planet, it would have increased its semi-major
axis by a factor $a/a_{\rm init} = M_{\rm prog}/M \sim 3.3$ as the
host adiabatically expelled its envelope.
We find that, for $M = 0.6\,M_\odot$, the orbital semi-major axis
is fairly tightly constrained to $a=2.3\pm 0.3 \,$AU, implying
$a_{\rm init}=0.7\pm0.1\,$AU.   It is unlikely that such a close
planet would survive the AGB phase of stellar evolution.  Of course,
a white dwarf need not be right at the peak.  For lower mass progenitors,
the ratio of initial to final masses is lower, which would enhance
the probability of survival.  But it is also the case that such white
dwarfs are rarer.

\subsection{Physical consistency checks of bayesian analysis}

The results reported here have been derived with the aid of
fairly complicated machinery, both in fitting the light curves
and in transforming from microlensing to physical parameters.
In particular, we have identified a strong mathematical degeneracy
between the parameters $\pi_{\rm E,N}$ and $\omega$, which
arise from orbital motion of the Earth and the planet, respectively.
When considering ``MCMC-only'' solutions,
this degeneracy led to extremely large errors in  $\pi_{\rm E,N}$
in Figure~\ref{fig:triangle}, which are then reflected in similarly
large errors in the ``light-curve-only'' contours for host mass and
lens-source relative parallax in Figure~\ref{fig:mpirel}.
Nevertheless, these large errors gradually shrink when the priors are applied
in Figure~\ref{fig:mpirel}, and more so when the constraints from
VLT observations are added in Figure~\ref{fig:mass-alone}.

We have emphasized that the high-$\pi_{\rm E}$ (so low-$D_L$, low-$M$)
solutions are very strongly, and improperly, favored by the MCMC when it
is cast in microlensing parameters, and that the Galactic prior
(Eq.~\ref{eqn:gamma}) properly compensates for this.  But is this
really true?  The best-fit distance for the Galactic-prior model
is four times larger than for the MCMC-only model, meaning that
the term $D_{\rm L}^4/\pi_{\rm rel}$ favors the Galactic model by a factor
$\sim 2500$.  Thus, even if the light curve strongly favored the nearby
model, the Galactic prior could ``trump'' the
light curve and enforce a larger distance.  
Indeed, this would be an issue if the Galactic prior were operating
by itself.
In fact, however,
Figure~\ref{fig:mpirel} shows that the finally adopted solution
(including the VLT flux constraint)
is disfavored by the light curve alone by just $\Delta\chi^2\sim 3$, so, in the end there is no strong tension.

A second issue is that both parallax and orbital motion are fairly
subtle effects that could, in principle, be affected by systematics.
If this were the case, the principal lensing parameters, such as
$q$ and $s$, would remain secure, but most of the ``higher order''
information, such as lens mass, distance, and orbital motion
would be compromised.  It is always difficult to test for systematics, particularly in this case for which 
there are two effects that are degenerate with each other
and in combination are detected at only $\Delta\chi^2<100$. 
}

\onecolumn{

 \begin{table}[h]
 \begin{center}
 \begin{tabular}{l r r r r r r r}
 
 \multicolumn{8}{c}{Table 3. Physical parameters}\\
 \hline
 Model & $M$ & $m_p$ & $D_L$ & $E_{\rm kin}/E_{\rm pot}$& $P$ & $a$ & $i$\\
 & $M_\odot$ & $M_{\rm Jup}$ & kpc &   & yr & AU & deg\\
 \hline
Kepler   & 0.04 &  0.51 &  2.29 &  0.34 &  2.92 &  1.39 &  39\\
90\% conf& $( 0.01, 0.12)$& $( 0.19, 1.69)$& $( 0.98, 4.79)$& $( 0.07, 0.44)$& $( 1.37, 5.42)$& $( 0.18, 2.10)$ & $( 24, 74)$\\
 \hline
Galactic & 0.31 &  4.38 &  6.83 &  0.54 &  3.73 &  2.12 &  60\\
90\% conf& $( 0.07, 6.37)$& $( 1.03,89.61)$& $( 3.65, 9.37)$& $( 0.06, 1.81)$& $( 1.37, 6.26)$& $( 1.06, 3.01)$ & $( 40, 79)$\\
 \hline
Gal+Kep  & 0.28 &  3.82 &  6.44 &  0.28 &  4.99 &  2.04 &  50\\
90\% conf& $( 0.07, 2.22)$& $( 1.00,30.82)$& $( 3.59, 9.38)$& $( 0.09, 0.37)$& $( 2.68, 7.27)$& $( 1.11, 3.04)$ & $( 38, 72)$\\
 \hline
Gal+VLT  & 0.25 &  3.55 &  6.42 &  0.69 &  4.90 &  1.62 &  58\\
90\% conf& $( 0.07, 0.53)$& $( 1.04, 7.52)$& $( 3.62, 8.34)$& $( 0.12, 1.99)$& $( 3.50, 6.79)$& $( 0.98, 2.45)$ & $( 42, 84)$\\
 \hline
G+K+VLT  & 0.19 &  2.56 &  5.69 &  0.27 &  5.43 &  1.82 &  52\\
90\% conf& $( 0.07, 0.49)$& $( 0.98, 6.71)$& $( 3.50, 7.87)$& $( 0.10, 0.36)$& $( 3.82, 7.58)$& $( 1.09, 2.68)$ & $( 40, 72)$\\
 \hline
 \end{tabular}
 \end{center}
 \end{table}

\begin{figure}[h!]
\centering
\includegraphics[width=6in]{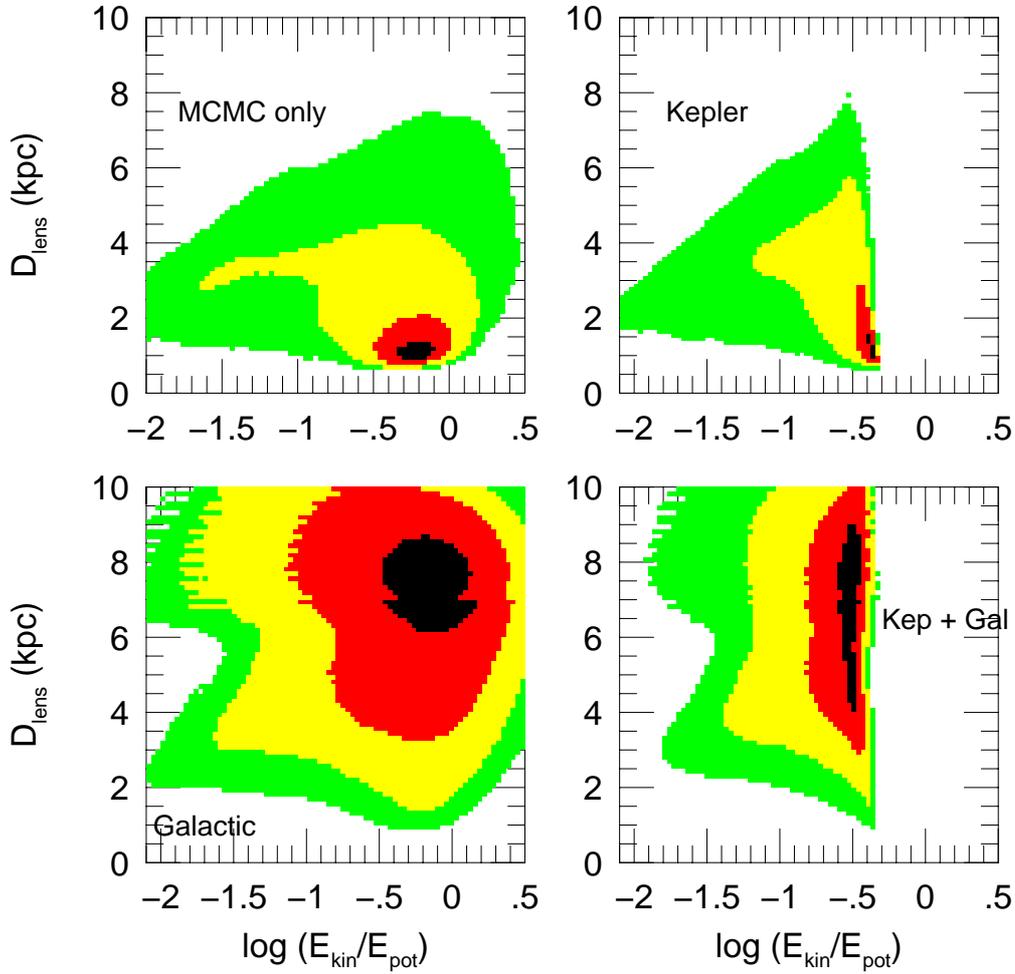}
\caption{\label{fig:betatest} Physical test of Bayesian
results: physicality diagnostic $\beta=E_{{\rm kin},\perp}/E_{{\rm pot},\perp}$
is plotted against host distance.  Bound orbits {\it must} have $\beta<1$,
and we expect a priori 
$0.1< \beta < 0.5$.}
\end{figure}

}

\twocolumn{

However, we can in fact test for such systematics using the diagnostic
\begin{equation}
\beta\equiv {v_\perp^2\over v_{{\rm esc},\perp}^2} = 
{E_{{\rm kin},\perp}\over E_{{\rm pot},\perp}},
\label{eqn:betadef}
\end{equation}
where $v_\perp$ and $v_{{\rm esc},\perp}$ are defined in Eqs.
(\ref{eqn:vperp}) and (\ref{eqn:vescperp}).  Bound orbits require
$\beta<1$.  Circular orbits, if seen face-on, have $\beta=0.5$ and
otherwise $\beta<0.5$.  Of course, it is possible to have 
$\beta\ll 1$, but it requires very special configurations to achieve
this.  For example, if the planet is close to transiting its host,
or if the orbit is edge-on and the phase is near quadrature.
Thus, a clear signature of systematics would be $\beta>1$ for all
light-curve solutions with reasonable $\chi^2$.  And if $\beta\la 0.1$,
one should be concerned about systematics, although this condition
would certainly not be proof of systematics.  With these considerations
in mind, we plot $D_{\rm L}$ vs.\ $\beta$ in Figure~\ref{fig:betatest}.

The key point is that the $1\,\sigma$ region of the Galactic-prior
panel straddles the region $\beta\la 0.5$ ($\log\beta\la -0.7$),
which is characteristic of approximately circular, approximately
face-on orbits.  It is important to emphasize that no selection
or weighting by orbital characteristics has gone into construction
of this panel.  This is a test which could easily have been failed
if the orbital parameters were seriously influenced by systematics:
$\beta$ could have taken literally on any value.

Finally, we turn to the two righthand panels, which incorporate
the orbital constraints.  Since these assume circular orbits,
they naturally eliminate all solutions with $\beta>0.5$, and some
smaller-$\beta$ solutions as well, because when $ds/dt\not=0$, it
is impossible to accommodate a $\beta=0.5$ circular orbit.  While
this radical censoring of the high-$\beta$ solutions is the most dramatic
aspect of these plots, there is also the very interesting effect that
low-$\beta$ solutions are also suppressed (though more gently).
This is because, as mentioned above, these require special configurations
and so are disfavored by the Kepler Jacobian, Eq.~(\ref{eqn:jac}).
Of course, radical censorship of $\beta>1$ solutions is entirely
appropriate (provided that $\beta<1$ solutions exist at reasonable
$\chi^2$), but what about $0.5\la\beta<1$?  A more sophisticated
approach would permit non-circular orbits and then suppress
these solutions ``more gently'' using a Jacobian (as is already
being for done low-$\beta$ solutions).  However, as we have emphasized,
the limited sensitivity of this event to additional orbital parameters
does not warrant such an approach.  Hence, radical truncation
is a reasonable proxy in the present case for the ``gentler'' and 
more sophisticated approach.

Moreover, one can see by comparing Rows 2 and 3 of Table 3
that the
addition of Kepler priors does not markedly alter the
Galactic-prior solutions.

\section{Conclusions}

We report the discovery of the planetary event MOA-2009-BLG-387Lb.
The planet/star mass ratio is very well-determined, $q=0.0132\pm 0.0003$.
We constrain the host mass to lie in the interval.
$0.07<M_{\rm host}/M_\odot<0.49$ at 90\% confidence, which corresponds to 
the full range of M dwarfs.  The planet mass therefore lies in
the range $1.0<m_p/M_{\rm Jup}<6.7$ , with its uncertainty
almost entirely due to the uncertainty in the host mass.
The host mass is determined from two ``higher-order''
microlensing parameters, $\theta_{\rm E}$ and $\bpi_{\rm E}$,
(i.e., $M=\theta_{\rm E}/\kappa\pi_{\rm E}$).

The first of these, the angular Einstein radius is actually quite
well measured, $\theta_{\rm E} = 0.31\pm 0.03\,$mas, from four separate
caustic-crossings by the source during the event. On the other
hand, from the light-curve analysis alone, the microlensing parallax
vector $\bpi_{\rm E}$ is poorly constrained because one of its
components is degenerate with a parameter describing orbital motion
of the lens.  That is, effects of the orbital motion of our planet
(Earth) and the lens planet have a similar impact on the light curve
and are difficult to disentangle.

Nevertheless, the closest-lens (and so also lowest-lens-mass)
solutions permitted by the light curve
are strongly disfavored by the Galactic model simply because there
are relatively few extreme-foreground lenses that can reproduce the
observed light-curve parameters.  Of course, we cannot
absolutely rule out the possibility that we are victims of chance,
so in principle it is possible that the host is an extremely
low-mass brown dwarf, or even a planet, with a lunar companion.

On the other hand, the arguments against a higher mass lens rest
on directly observed features of the light curve.  That is, 
as mentioned above, $\theta_{\rm E}$ is measured accurately from
the four observed caustic crossings.  And one component of $\bpi_{\rm E}$,
the one in the projected direction of the Sun, is also 
reasonably well measured from the observed asymmetry in the light curve outside the
caustic region.  This places a lower limit on $\pi_{\rm E}$,
hence an upper limit on the mass.

However, for the latter parameter, the very strong prior from the 
Galactic model favoring more distant lenses would, by itself,
``overpower'' the lightcurve and impose solutions with
$M>1\,M_\odot$, which are disfavored by the lightcurve at
$>3\,\sigma$.  It is only because these high-mass solutions
are ruled out by flux limits from VLT imaging that the 
lightcurve-only $\chi^2$ is quite compatible with the final,
posterior-probability solution.

The relatively high planet/star mass
ratio (implying a Jupiter-mass planet for the case of
a very late M-dwarf host)
is then difficult to explain within the context of the standard
core-accretion paradigm.

The 12-day duration of the planetary perturbation, one of the longest
seen for a planetary microlensing event, enabled us to detect
two components of the orbital motion, basically the projected velocity
in the plane of the sky perpendicular and parallel to the star-planet
separation vector.  While the first of these is strongly degenerate
with the microlens parallax (as mentioned above), the second one
(which induces a changing shape of the caustic) is reasonably well
constrained by the two sets of well-separated double caustic crossings.
Moreover, once the Galactic-model prior constrained the microlensing
parallax, its correlated orbital parameter was implicitly constrained
as well.  With two orbital parameters, plus two position parameters
from the basic microlensing fit (projected separation $s$, and orientation
of the binary axis relative to the source motion $\alpha$) plus the
lens mass, there is enough information to specify an orbit, if the
orbit is assumed circular.  We are thus able to estimate
a semi-major axis $a=1.8\,$AU and period $5.4\,$ years.

We recognized that inferences derived from such subtle light curve
effects could in principle be compromised by systematics.  We therefore
tested whether the derived ratio of orbital kinetic to potential
energy was in the expected range, before imposing any orbital
constraints.  If the measurements were strongly influenced by 
systematic errors, this ratio could have taken on any value.
In fact, it fell right in the expected range.

\begin{acknowledgements}
Based on observations with the European Southern Observatory telescopes from the ESO/ST-ECF Science Active Facility, Chile (short programme, run 385.C-0797).
VB thanks Ohio State University for its hospitality during a six week visit, during which this study was initiated. 
We acknowledge the following support:
Grants HOLMES ANR-06-BLAN-0416
Dave Warren for the Mt Canopus Observatory;
NSF AST-0757888 (AG,SD); NASA NNG04GL51G (DD,AG,RP);
Polish MNiSW N20303032/4275 (AU);
HST-GO-11311 (KS);
NSF AST-0206189 and AST-0708890, NASA NAF5-13042 and NNX07AL71G (DPB);
Korea Science and Engineering Foundation grant 2009-008561 (CH);
Korea Research Foundation grant 2006-311-C00072 (B-GP);
Korea Astronomy and Space Science Institute (KASI);
Deutsche Forschungsgemeinschaft (CSB);
PPARC/STFC, EU FP6 programme ``ANGLES'' ({\L}W,NJR);
PPARC/STFC (RoboNet);
Dill Faulkes Educational Trust (Faulkes Telescope North);
Grants JSPS18253002, JSPS20340052 and JSPS19340058 (MOA);
Marsden Fund of NZ(IAB, PCMY); 
Foundation for Research Science and Technology of NZ;
Creative Research Initiative program (2009-008561) (CH);
Grants MEXT19015005 and JSPS18749004 (TS).
Work by SD was performed under contract with the California Institute
of Technology (Caltech) funded by NASA through the Sagan Fellowship
Program.
JCY is supported by an NSF Graduate Research Fellowship.
This work was supported in part by an allocation of computing time from the Ohio Supercomputer Center.
JA is supported by the Chinese Academy of Sciences (CAS) Fellowships for Young International Scientist, Grant No.:2009Y2AJ7.
\end{acknowledgements}

}

\twocolumn{



}
}
\end{document}